\documentclass{aa}
\usepackage{graphicx}
\usepackage{times}
\usepackage{url}
\newcommand{\EQ}{\begin{equation}}
\newcommand{\EN}{\end{equation}}
\newcommand{\EQA}{\begin{eqnarray}}
\newcommand{\ENA}{\end{eqnarray}}

\newcommand{\Eq}[1]{Eq.~(\ref{#1})}

\newcommand{\Eqss}[2]{Eqs~(\ref{#1})--(\ref{#2})}
\newcommand{\eqs}[2]{(\ref{#1}) and~(\ref{#2})}

\newcommand{\Sec}[1]{Sect.~\ref{#1}}

\newcommand{\Fig}[1]{Fig.~\ref{#1}}
\newcommand{\FFig}[1]{Figure~\ref{#1}}
\newcommand{\Tab}[1]{Table~\ref{#1}}
\newcommand{\Figs}[2]{Figs~\ref{#1} and \ref{#2}}

\newcommand{\bra}[1]{\langle #1\rangle}

\newcommand{\meanEMF}{\overline{\vec{\cal E}}}

\newcommand{\meanB}{\overline{B}}

\newcommand{\meanBB}{\overline{\vec{B}}}
\newcommand{\meanJJ}{\overline{\vec{J}}}
\newcommand{\meanUU}{\overline{\vec{U}}}
\newcommand{\meanWW}{\overline{\vec{W}}}
\newcommand{\meanTT}{\overline{\vec{T}}}

\newcommand{\eee}{\hat{\mbox{\boldmath $e$}} {}}

\newcommand{\xx}{\mbox{\boldmath $x$} {}}

\newcommand{\UU}{{\vec{U}}}
\newcommand{\uu}{{\vec{u}}}
\newcommand{\BB}{{\vec{B}}}
\newcommand{\JJ}{{\vec{J}}}
\newcommand{\jj}{{\vec{j}}}

\newcommand{\aaaa}{{\vec{a}}}
\newcommand{\bb}{{\vec{b}}}

\newcommand{\ff}{\mbox{\boldmath $f$} {}}

\newcommand{\FF}{{\vec{F}}}
\newcommand{\TT}{\mbox{\boldmath $T$} {}}

\newcommand{\kk}{\mbox{\boldmath $k$} {}}

\newcommand{\nab}{\mbox{\boldmath $\nabla$} {}}

\newcommand{\oo}{\mbox{\boldmath $\omega$} {}}

\newcommand{\RRRR}{\mbox{\boldmath ${\sf R}$} {}}
\newcommand{\ii}{{\rm i}}

\newcommand{\dd}{{\rm d} {}}
\newcommand{\const}{{\rm const}  {}}

\def\half{{\textstyle{1\over2}}}

\def\onethird{{\textstyle{1\over3}}}

\newcommand{\yapj}[3]{ #1, {ApJ,} {#2}, #3}

\newcommand{\yan}[3]{ #1, {AN,} {#2}, #3}

\newcommand{\yana}[3]{ #1, {A\&A,} {#2}, #3}

\newcommand{\ygafd}[3]{ #1, {GApFD,} {#2}, #3}
\newcommand{\yjfm}[3]{ #1, {JFM,} {#2}, #3}
\newcommand{\ypf}[3]{ #1, {PhFl,} {#2}, #3}
\newcommand{\ypp}[3]{ #1, {Phys. Plasmas,} {#2}, #3}

\newcommand{\yjetp}[3]{ #1, {Sov. Phys. JETP,} {#2}, #3}

\newcommand{\yprl}[3]{ #1, {PRL,} {#2}, #3}
\newcommand{\ypre}[3]{ #1, {PRE,} {#2}, #3}

\newcommand{\ymn}[3]{ #1, {MNRAS,} {#2}, #3}

\newcommand{\pprt}[2]{ #1, {PhR, }{#2} (in press)}

\newcommand{\yjour}[4]{ #1, {#2}, {#3}, #4}

\newcommand{\ybook}[3]{ #1, {#2} (#3)}

\begin{document}

\titlerunning{Minimal tau approximation}
\title{Minimal tau approximation and simulations of the alpha effect}
\author{A.\ Brandenburg\inst{1} \and K.\ Subramanian\inst{2}}

\institute{
NORDITA, Blegdamsvej 17, DK-2100 Copenhagen \O, Denmark
\and
IUCAA, Post Bag 4, Pune University Campus, Ganeshkhind, Pune 411 007, India
}

\date{Received 10 April 2005 / Accepted 30 May 2005}

\abstract{
The validity of a closure called
the minimal tau approximation (MTA), 
is tested in the context of dynamo theory, 
wherein triple correlations are assumed to 
provide relaxation of the turbulent electromotive force.
Under MTA, the alpha effect in mean field dynamo theory becomes
proportional to a relaxation time scale multiplied by
the difference between kinetic and current helicities.
It is shown that the value of the relaxation time is positive and,
in units of the turnover time at the forcing wavenumber,
it is of the order of unity.
It is quenched by the magnetic field -- roughly independently of the
magnetic Reynolds number.
However, this independence becomes uncertain
at large magnetic Reynolds number.
Kinetic and current helicities are shown to be dominated by
large scale properties of the flow.
 }
\maketitle

\section{Introduction}

In many branches of astrophysics it is necessary to use turbulent transport
coefficients.
This allows one to describe the collective effects of turbulence
(e.g.\ diffusive and non-diffusive effects) without the need to resolve
the small scale turbulence.
Non-diffusive effects include the $\alpha$ effect in mean-field dynamo
theory (Krause \& R\"adler 1989) and the $\Lambda$ effect in the theory
of stellar differential rotation (R\"udiger 1989).
In order to calculate turbulent transport coefficients
one usually makes the assumption that the
equations for the fluctuating quantities can be linearized
(Moffatt 1978, Krause \& R\"adler 1980).
This implies that one is forced to restrict oneself to the case where
the fluctuations are weak compared with the mean field.
This is clearly not very useful for astrophysical applications.
Nevertheless, mean-field theory is usually applied in the astrophysically
interesting parameter regime (e.g., R\"udiger \& Kitchatinov 1993,
Kitchatinov et al.\ 1994), which should in principle be well beyond the
regime of validity of such a linear approximation.
How is it then possible that the results are actually quite reasonable?

The approximation under which higher order terms are neglected is known
as the quasilinear, the first order smoothing, or the second order
correlation approximation.
With the exception on an important additional term,
the results derived under this approximation turn out to be similar to
those derived under a more general approximation, which is sometimes
referred to as the minimal tau approximation (MTA).
This approximation has its roots
in early papers by Vainshtein \& Kitchatinov (1983),
Kleeorin et al.\ (1990, 1996),
and its general usefulness has recently been stressed further
with the papers by Blackman \& Field (2002) and R\"adler et al.\ (2003).
A recent review of nonlinear dynamo theory and MTA has 
been given by Brandenburg \& Subramanian (2005a).
MTA has also been applied to the case of passive scalar diffusion
(Blackman \& Field 2003), where its validity has been tested numerically
(Brandenburg et al.\ 2004).

In the hydromagnetic case, the single most important addition to the
theory is the attenuation of the $\alpha$ effect by a term proportional
to the current helicity of the small scale field.
The beauty of MTA is that this term emerges in a completely natural fashion.
In the first order smoothing approach this term is absent, although even
there it can be incorporated in a more phenomenological and hence less
convincing fashion (Brandenburg \& Subramanian 2005a).
The role of the current helicity term was first identified by
Pouquet et al.\ (1976) and later associated with catastrophic
quenching by Gruzinov \& Diamond (1994) and many others after them.
The purpose of the present paper is to test MTA in the context of
dynamo theory.

\section{The method}

In mean field dynamo theory one splits the magnetic
field $\BB$ into a mean 
magnetic field $\meanBB$ and a small scale
field  $\bb =\BB - \meanBB$, and 
derives the mean-field dynamo equation
\EQ
{\partial\meanBB\over\partial t}=
\nab\times\left(\meanUU\times\meanBB+\meanEMF-\eta\meanJJ\right).
\label{basicm}
\EN
Here $\meanEMF\equiv\overline{\uu\times\bb}$ is the turbulent
electromotive force, $\meanJJ=\nab\times\meanBB/\mu_0$
the mean current density, $\mu_0$ the vacuum permeability
(assumed unity throughout the rest of the paper),
and the velocity $\UU = \meanUU + \uu$ has also
been split into mean $\meanUU$ and small scale
$\uu = \UU - \meanUU$ velocities.
The crucial difference between the first order smoothing approximation
and MTA is that, instead of writing an expression for the electromotive
force $\meanEMF\equiv\overline{\uu\times\bb}$, one considers the expression
for its time derivative,
\EQ
{\partial\meanEMF\over\partial t}
=\overline{\dot\uu\times\bb}
+\overline{\uu\times\dot\bb}.
\EN
The evolution equations for $\dot\uu$ and $\dot\bb$ contain terms that
are linear in the fluctuations.
These terms lead to quadratic correlations of velocity (which are considered
to be known from a turbulence model) and quadratic correlations
of the magnetic field (which have
to be obtained from a dynamical feedback model).
In addition, there are triple moments resulting from the quadratic
nonlinearities in the equations for $\dot\uu$ and $\dot\bb$.
The sum of all triple moments is denoted by $\meanTT$ and is assumed to
be equal to the normalized quadratic correlations $-\meanEMF/\tau$, i.e.\
\EQ
\meanTT=-{\meanEMF\over\tau}\quad\mbox{(closure hypothesis)}.
\label{closure}
\EN
Blackman \& Field (2002) motivate this term on physical grounds: in
the absence of any helical ``driving'', i.e.\ when the quadratic
correlations vanish, $\meanEMF$ should always decay.
It is mainly the numerical simulations that can lend some support to
this otherwise rather ad hoc and ill-justified assumption.
[In principle one can also do this closure in Fourier space
and adopt a $k$-dependent $\tau(k)$, as has been done by
R\"adler et al.\ (2003) for example; this is discussed
in more detail in Brandenburg \& Subramanian (2005a).]

As was first shown by Blackman \& Field (2002), the use of MTA leads to
an explicitly time-dependent equation for the electromotive force which,
in the isotropic approximation, is
\EQ
{\partial\meanEMF\over\partial t}=\tilde\alpha\meanBB
-\tilde\eta_{\rm t}\meanJJ-{\meanEMF\over\tau}\quad\mbox{(MTA)}.
\label{dEMFdt}
\EN
This has to be contrasted with the usual result from the first order
smoothing approximation (FOSA),
\EQ
\meanEMF=\alpha\meanBB-\eta_{\rm t}\meanJJ\quad\mbox{(FOSA)},
\EN
which agrees with \Eq{dEMFdt} if the time derivative of $\meanEMF$ is
neglected.
Here, $\alpha=\tau\tilde\alpha$ is the usual $\alpha$ effect and
$\eta_{\rm t}=\tau\tilde\eta_{\rm t}$ is the turbulent magnetic diffusivity.
One should point out, however, that the meaning of $\tau$ is different in
the two cases, and that FOSA can only be justified if the $\tau$
in this formalism is small.

In the passive scalar case a relatively robust method
for estimating $\tau$ from turbulence simulations
is to impose a mean concentration gradient across
the box, and to treat the deviations from this mean gradient as periodic
in all three directions (Brandenburg et al.\ 2004).
In the present case this corresponds to imposing a mean magnetic field,
$\BB_0=\mbox{constant}$.
The magnetic field is then given by $\BB=\BB_0+\bb$, where $\bb$
is the departure from the mean field, $\bb$.
Since $\partial\BB_0/\partial t=0$, the evolution of $\bb$ is given by
\EQ
\dot{\bb}=\nab\times
[\uu\times(\BB_0+\bb)-\eta\jj+\ff_{\rm mag}],
\label{bdot}
\EN
where a dot denotes a time derivative,
$\jj=\nab\times\bb$ is the current density,
and the current density is measured in units were $\mu_0=1$.
We have allowed for the possibility of a magnetic forcing term,
$\ff_{\rm mag}$, which is set to zero unless stated otherwise.
We assume that the gas is weakly compressible and isothermal with constant
sound speed $c_{\rm s}$.
The evolution equation for the velocity is therefore
\EQ
\dot{\uu}=
-\uu\cdot\nab\uu-\nab h+\jj\times(\BB_0+\bb)/\rho
+\ff_{\rm kin}+\FF_{\rm visc},
\label{udot}
\EN
where $h$ is proportional to the enthalpy.
Since the ratio of specific heats is unity, we have
$h=h_0+c_{\rm s}\ln\rho$, where $h_0$ is a constant
whose value is unimportant for the dynamics.
The evolution equation of $\ln\rho$ is
\EQ
\partial\ln\rho\,/\,\partial t=-\uu\cdot\nab\ln\rho-\nab\cdot\uu.
\label{lrdot}
\EN

We define a generic forcing function $\ff$
(see Appendix~\ref{ForcingFunction}) and put $\ff_{\rm kin}=\ff$ in
the kinetically driven case ($\ff_{\rm mag}=0$), or $\ff_{\rm mag}=\ff$
with $\ff_{\rm kin}=0$ in the magnetically driven case.
The wavevector $\kk(t)$ of the forcing function is delta-correlated
in time, so at each timestep a new vector is chosen randomly with
$k_{\rm f}-\half\leq|\kk|\leq k_{\rm f}+\half$.
We consider two cases: $k_{\rm f}=1.5$ and $k_{\rm f}=5$.
The forcing function is chosen to be maximally helical with
positive helicity.

In the main part of this paper we adopt combined volume and
time averages, so, $\meanBB=\BB_0=\const$.
This means that in \Eq{dEMFdt} both $\meanJJ$ and
$\partial\meanEMF/\partial t$ vanish.
Therefore we simply have
\EQ
\meanEMF=\tau\tilde\alpha\BB_0.
\EN
Expressing $\meanEMF$ in terms of $\alpha=\bra{\meanEMF\cdot\BB_0}_t/\BB_0^2$,
where $\bra{...}_t$ denotes a combined volume and time average,
we have simply $\alpha=\tau\tilde\alpha$.
Here, $\tilde\alpha$ only depends on the quadratic correlations.
Using the evolution equations \eqs{bdot}{udot}, this leads to
\EQ
\tilde\alpha=\onethird
\left(-\overline{\oo\cdot\uu}
+\rho_0^{-1}\overline{\jj\cdot\bb}\right).
\label{alpiso}
\EN
This important relation was first obtained by Pouquet et al.\ (1976).
In the following we will use the abbreviations
\EQ
\tilde\alpha=\tilde\alpha_{\rm K}+\tilde\alpha_{\rm M},\quad
\tilde\alpha_{\rm K}=-\onethird\overline{\oo\cdot\uu},\quad
\tilde\alpha_{\rm M}=\onethird\rho_0^{-1}\overline{\jj\cdot\bb}.
\label{alphaKM}
\EN
The triple moment, $\meanTT$,
has two separate contributions from the two evolution equations,
$\meanTT=\meanTT_{\rm M}+\meanTT_{\rm K}$, where
\EQ
\meanTT_{\rm M}=\overline{\uu\times\nab\times(\uu\times\bb)},
\EN
\EQ
\meanTT_{\rm K}=
\overline{(-\uu\cdot\nab\uu-\nab h+\jj\times\tilde\bb/\rho)\times\bb},
\EN
and $\tilde\bb=\bb+\BB_0(1-\rho/\rho_0)$ includes the fluctuating
contribution from the applied field due to density fluctuations.

So far, and throughout most of this paper,
we consider the case of an imposed constant mean field, $\BB_0$,
with zero mean current, $\JJ_0=\nab\times\BB_0=0$.
While this is reasonable for measuring the value of $\alpha$,
it is of course unrealistic for dynamo calculations, where there
must be a mean current proportional to the curl of the mean field.
Furthermore, if a mean field is imposed, this mean field
can obviously not change by the dynamo action.
Nevertheless, it provides a useful method for calculating $\alpha$.
This method gives similar results than the more complicated and less accurate
methods that can be invoked in dynamo simulations (Brandenburg 2001,
see his Figs~14 and 15).
However, in order to show that \Eq{closure} has any justification,
we insert a discussion (\Sec{LocalCorrel}) where we consider the local
correlation in a simulation without an imposed field.

\section{Spatial variation of $\TT$}
\label{LocalCorrel}

Before we present in detail results for the case of an imposed field, we
consider first the case of a dynamo-generated magnetic field.
In that case the magnetic field shows a marked sinusoidal variation
in one of three possible directions.
Which of the three directions is chosen depends on chance.
In the following we compare with a simulation of homogeneous
helically forced turbulence of Brandenburg (2001); in his Run~3
the magnetic field shows a slow variation in the $y$ direction,
so we define mean fields by averaging over the $x$ and $z$ directions.

\begin{figure}[t!]\begin{center}
\includegraphics[width=\columnwidth]{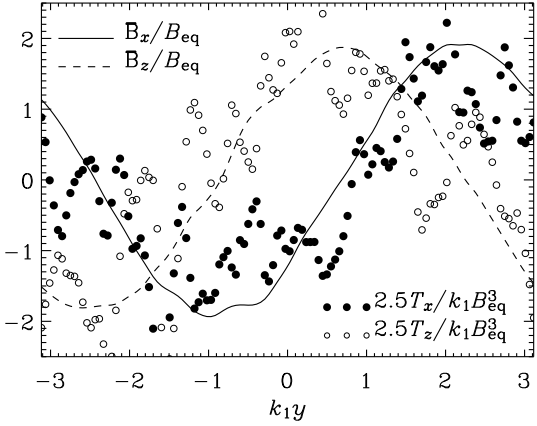}
\end{center}\caption[]{
Comparison of the spatial dependence of two components of the mean
magnetic field and the triple correlation in Run~3 of B01. 
The magnetic field is normalized by the equipartition field strength,
$B_{\rm eq}$, and the triple correlation is normalized by
$k_1B_{\rm eq}^3$, but scaled by a factor of $2.5$ to make it have a similar
amplitude as the mean field.
Note that $\meanB_x$ (solid line) correlates with $T_x$ (filled dots)
and $\meanB_z$ (dashed line) correlates with $T_z$ (open dots).
}\label{ptriple}\end{figure}

\FFig{ptriple} shows that the $x$ and $z$ components of $\meanTT$
have a positive correlation with the mean magnetic field.
Note that $\meanEMF$ itself has a negative correlation with $\meanBB$,
for a negative $\alpha$. And since this simulation has positive 
kinetic helicity, the $\alpha$ obtained is indeed negative.
Therefore the positive correlation
of $\meanTT$ with $\meanBB$ 
is consistent with \Eq{closure} with a positive $\tau$.
(The term $\eta_{\rm t}\meanJJ$ does not change the sign of
$\alpha\meanBB$, but only reduces its magnitude. This follows
from the fact that in a dynamo simulation $\meanEMF$ has to
overcome the magnetic diffusion term.)
We have used $\meanBB$ as a proxy for $\meanEMF$ in order to study the
correlation of $\meanTT$ with $\meanEMF$, because
$\meanEMF$ itself is even more noisy than $\meanTT(y)$.
This is because $\alpha\meanBB$ almost cancels
$(\eta+\eta_{\rm t})\meanJJ$ and the fact that $\meanJJ$ consists of
derivatives of $\meanBB$ which contributes further to the noise.

In conclusion, our first assessment of $\meanTT$ is very encouraging
in that its phase relation with the magnetic field is such that
it would indeed contribute as a damping term, i.e.\ that it has
the expected sign.

\section{Reynolds number dependence of $\alpha$ and $\tilde\alpha$}
\label{RmDependence}

Having shown that $\meanTT$ varies locally in the expected
sense, we now concentrate on the case of a uniform imposed field where
it is possible to use averages over the full box.
(In the dynamo case considered in the previous section such averages
would have given zero.)
We solve \Eqss{bdot}{lrdot} numerically.
The forcing amplitude is such that
we are in the low Mach number regime, i.e.\
$|\uu|\ll c_{\rm s}$, so for all practical purposes the flows can be
considered nearly incompressible.
We consider runs with different hydrodynamic and hydromagnetic Reynolds
numbers,
\EQ
\mbox{Re}=u_{\rm rms}/(\nu k_{\rm f}),\quad
R_{\rm m}=u_{\rm rms}/(\eta k_{\rm f}),
\label{defReRm}
\EN
but keep the magnetic Prandtl number unity in all cases, i.e.\
$R_{\rm m}=\mbox{Re}$.
Here, $k_{\rm f}$ is the wavenumber of the energy-carrying scale
which is assumed to be the forcing wavenumber of either 1.5 or 5.
The simulations have been carried out using the {\sc Pencil Code}\footnote{
\url{http://www.nordita.dk/software/pencil-code}} which is a high-order
finite-difference code (sixth order in space and third
order in time) for solving the compressible hydromagnetic equations.
In the following we consider four different values of $R_{\rm m}$ and
increase the resolution correspondingly by factors of 2 between $64^3$
and $512^3$ meshpoints,
and in some cases between $32^3$ and $256^3$ meshpoints.

\begin{figure}[t!]\begin{center}
\includegraphics[width=\columnwidth]{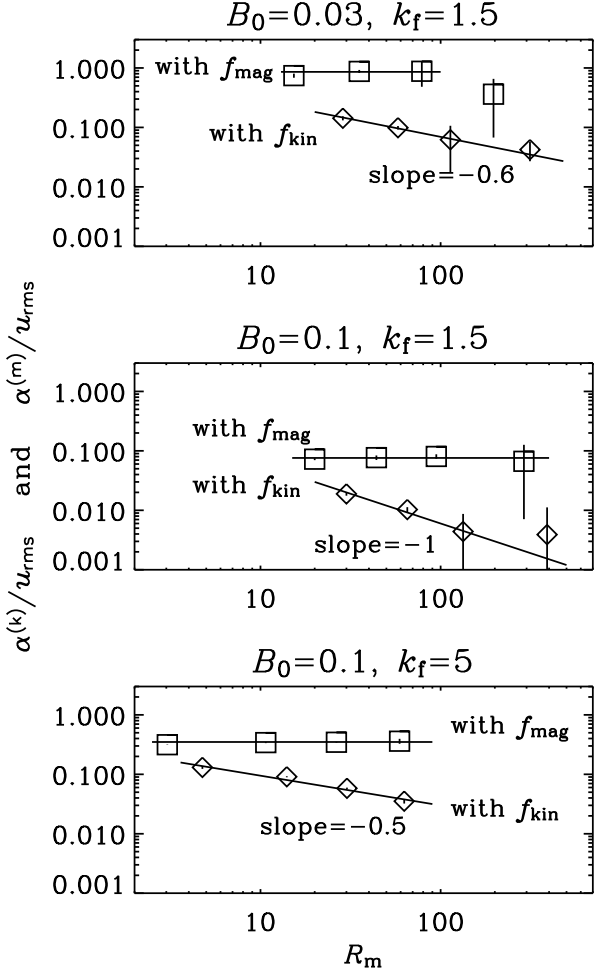}
\end{center}\caption[]{
$R_{\rm m}$ dependence of the normalized $\alpha$
for $B_0=0.03$ (or $B_0/B_{\rm eq}=0.3...0.6$) 
and $B_0=0.1$ (or $B_0/B_{\rm eq}=0.6...1.5$) in cases where the
turbulence is driven kinetically with a body force $\ff_{\rm kin}$
(lower graph) or magnetically with a forcing term $\ff_{\rm mag}$
in the induction equation (upper graph).
The vertical bars on the data points give estimates of the
error (see text).
}\label{palp_vs_Rm}\end{figure}

We measure velocity in units of the sound speed $c_{\rm s}$,
density in units of the average density $\rho_0$,
magnetic field in units of $\sqrt{\mu_0\rho_0}\,c_{\rm s}$,
and length in units of $k_1^{-1}$, where $k_1$ is the smallest
possible wavenumber in the box.
Whenever possible we present the result in normalized
(explicitly nondimensional) form.

We begin by considering the dependence of
$\alpha$ on $R_{\rm m}$; see Fig.~\ref{palp_vs_Rm}.
We focus on the results for two different field strengths,
$B_0=0.03$, corresponding to $B_0/B_{\rm eq}\approx0.45$, and
$B_0=0.1$, corresponding to $B_0/B_{\rm eq}\approx1$.
Throughout this paper, error bars are obtained by taking the extrema
from 5 separate averages, each over 1/5 of the full time series.

In both kinetically and magnetically driven cases $\alpha$ is
negative when the helicity of the forcing is positive.
It turns out that, when the turbulence is kinetically forced,
$-\alpha$ decreases with increasing $R_{\rm m}$ like $R_{\rm m}^{-n}$
where $n$ is between 1/2 and 1.
For large field strengths and large values of $R_{\rm m}$ we expect
an asymptotic dependence with $n=1$ (Cattaneo \& Hughes 1996),
but when these values are not yet
large enough the data are best fitted with values of $n$ less than 1.

In the magnetically driven case $\alpha$ is generally, at comparable
values of $u_{\rm rms}$, larger than in the kinetically driven case.
Furthermore, in the magnetically driven case $\alpha$ does not show the
systematic decrease of $R_{\rm m}$ seen in the kinetically driven case.
This can be understood as a result of a modified Keinigs (1983) relation;
see Appendix~\ref{Keinigs}.

\begin{figure}[t!]\begin{center}
\includegraphics[width=\columnwidth]{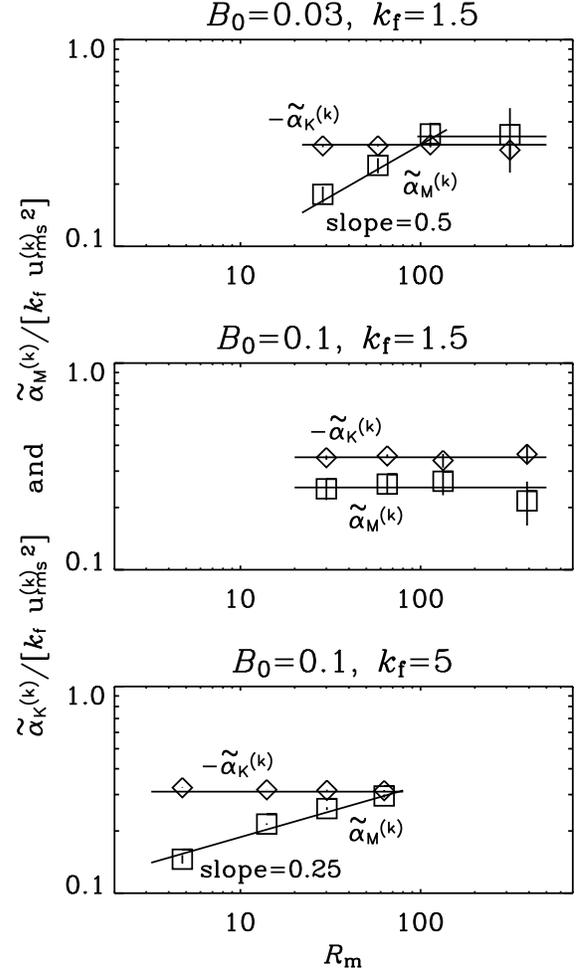}
\end{center}\caption[]{
Dependence of $\tilde{\alpha}_{\rm K}^{\rm(k)}$ and
$\tilde{\alpha}_{\rm M}^{\rm(k)}$
on $R_{\rm m}$ in the kinetically forced case.
Vertical bars give error estimates.
}\label{palpKMkin_vs_Rm}\end{figure}

Next, we consider the values of $\tilde{\alpha}_{\rm K}$ and
$\tilde{\alpha}_{\rm M}$ in both kinetically and magnetically
driven cases; see \Figs{palpKMkin_vs_Rm}{palpKMmag_vs_Rm}.
The results depend on the values of $B_{\rm 0}$ and $k_{\rm f}$.
In the kinetically driven case there is a clear tendency for
$\tilde{\alpha}_{\rm M}$ to increase with $R_{\rm m}$ until it
approaches and slightly exceeds the value of $-\tilde{\alpha}_{\rm K}$.
For larger values of $R_{\rm m}$, $-\tilde{\alpha}_{\rm K}$ and
$\tilde{\alpha}_{\rm M}$ are constant and close to each other.

\begin{figure}[t!]\begin{center}
\includegraphics[width=\columnwidth]{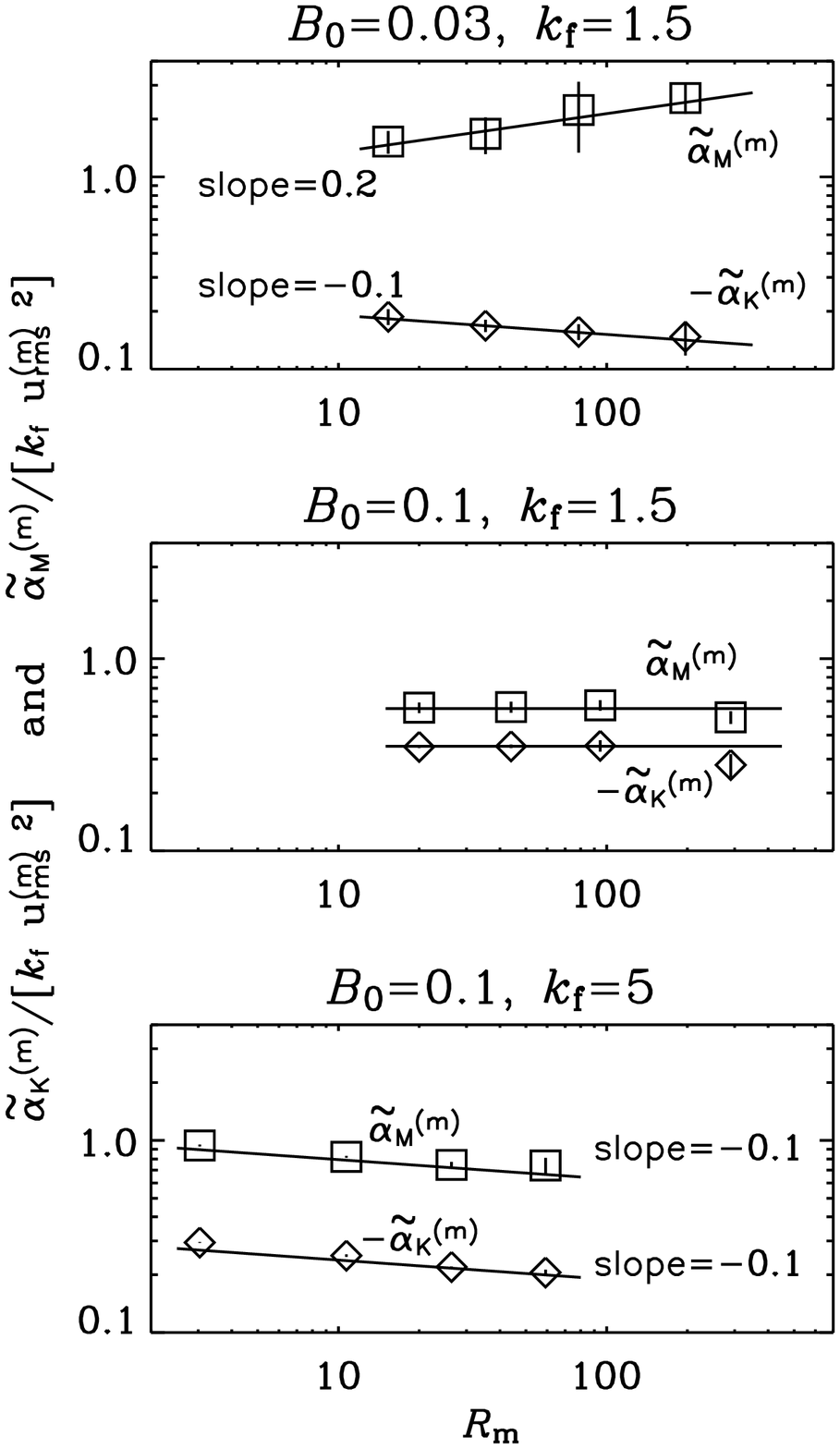}
\end{center}\caption[]{
Dependence of $\tilde{\alpha}_{\rm K}^{\rm(m)}$ and
$\tilde{\alpha}_{\rm M}^{\rm(m)}$
on $R_{\rm m}$ in the magnetically forced case.
Vertical bars give error estimates.
}\label{palpKMmag_vs_Rm}\end{figure}

In the magnetically driven case, $\tilde{\alpha}_{\rm K}$ and
$\tilde{\alpha}_{\rm M}$ depend only weakly on $R_{\rm m}$.
Furthermore, $\tilde{\alpha}_{\rm M}$ can exceed $-\tilde{\alpha}_{\rm K}$
by even an order of magnitude; see the first panel of \Fig{palpKMmag_vs_Rm}.

It is not easy to present the data of all runs in a meaningful
way in a single graph.
This is why we have summarized the results of all runs in tabular form;
see \Tab{Ttau}, allowing alternative representations and interpretations
to be made.

\section{The relaxation time}
\label{RelaxTime}

\begin{figure}[t!]\begin{center}
\includegraphics[width=\columnwidth]{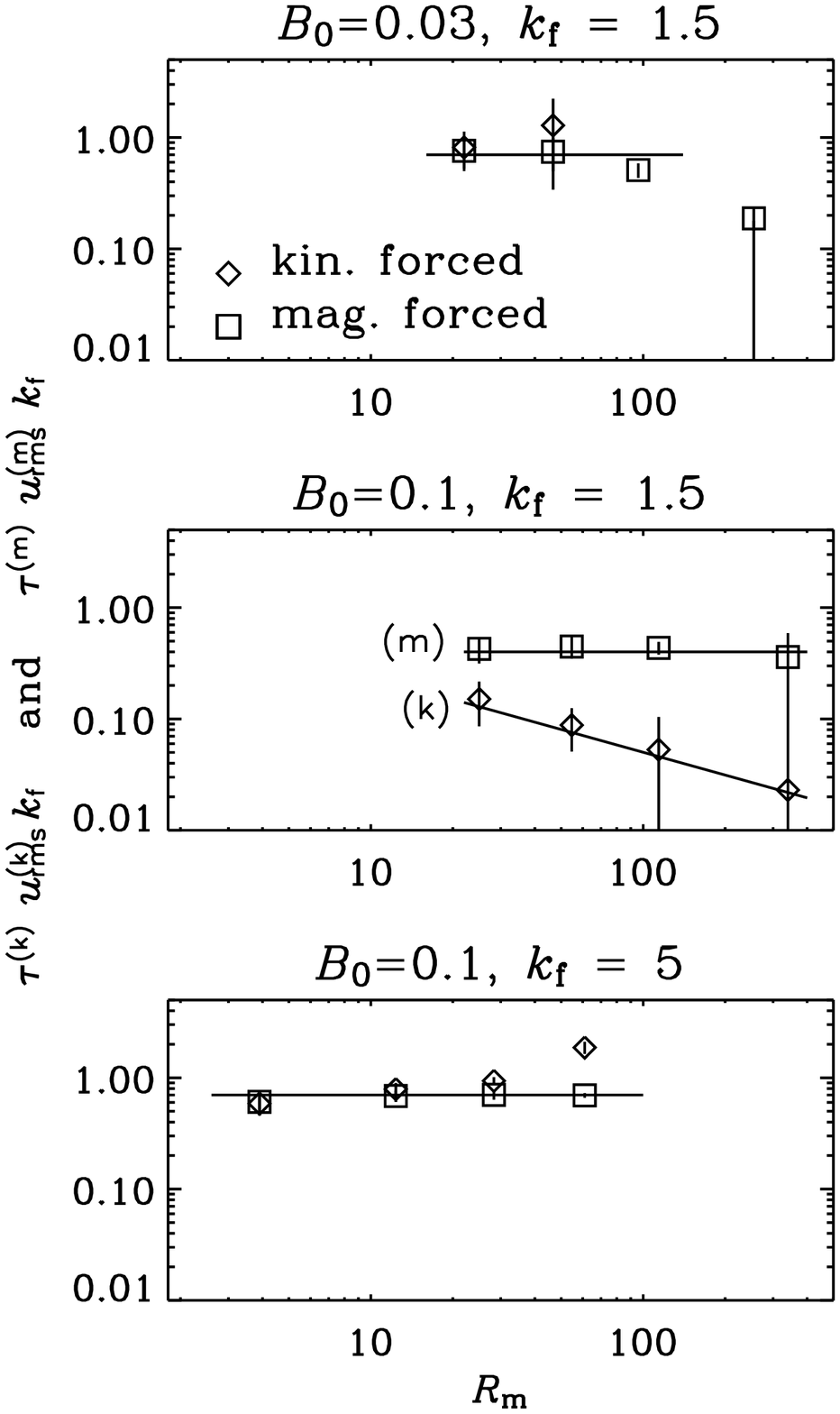}
\end{center}\caption[]{
Strouhal numbers as a function of $R_{\rm m}$
for different combinations of $B_0$ and $k_{\rm f}$.
Here, $g_{\rm K}=g_{\rm M}$ is assumed.
The horizontal lines are drawn to indicate the range over which
the Strouhal numbers are approximately constant.
In the second panel therefore, the slope of the lower graph is $-0.68$.
}\label{ptau1}\end{figure}

After these preparations we can now return to the main question addressed
in this paper: what is the relevant relaxation time in MTA and how does
it depend on $R_{\rm m}$.
Assuming a steady state, we can use \Eq{dEMFdt} and calculate
\EQ
\tau=\alpha/(\tilde{\alpha}_{\rm K}+\tilde{\alpha}_{\rm M}).
\EN
The value of $\tau$ obtained in this way
is shown in \Fig{ptau1}, where we show the results
separately for the kinetically and magnetically driven cases.
The general trend is quite similar to the dependence of $\alpha$
on $R_{\rm m}$.

\begin{figure}[t!]\begin{center}
\includegraphics[width=\columnwidth]{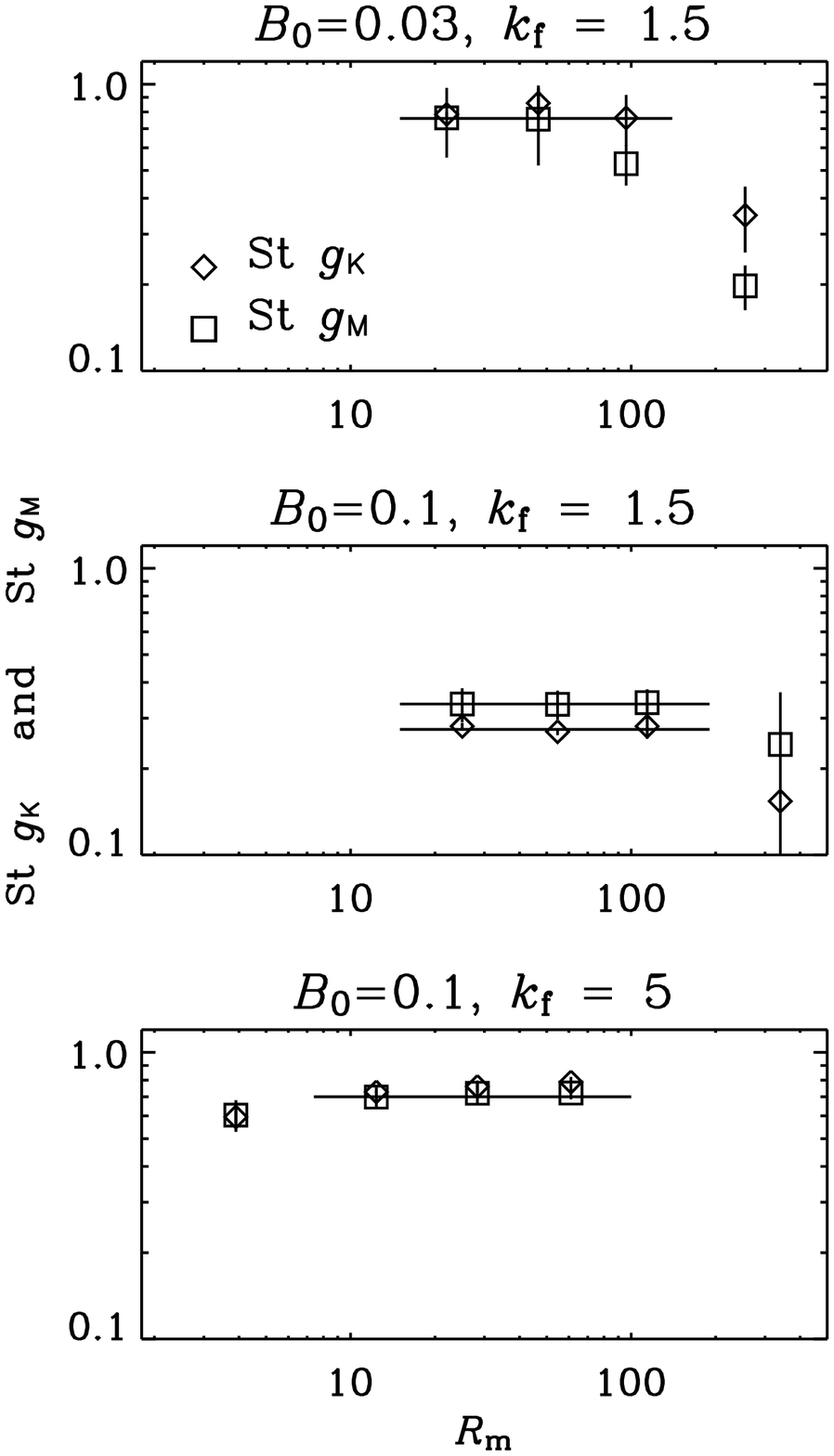}
\end{center}\caption[]{
Magnetic and kinetic Strouhal numbers as a function of $R_{\rm m}$
for different values of $B_0$ and  $k_{\rm f}$.
Here, kinetically and magnetically forced runs have been used to
calculate separately $g_{\rm K}\neq g_{\rm M}$.
The horizontal lines are drawn to indicate the range over which
the Strouhal numbers are approximately constant.
}\label{ptau}\end{figure}

As we have discussed above, at large $R_{\rm m}$,
$\tilde{\alpha}_{\rm M}$ can even slightly exceeds the value of
$-\tilde{\alpha}_{\rm K}$.
Since $\alpha$ itself does not change sign, this would imply that
$\tau$ would become negative for large $R_{\rm m}$.
An obvious solution to this problem is that $\tilde{\alpha}_{\rm K}$ and
$\tilde{\alpha}_{\rm M}$ should be preceded by some additional
quenching functions, $g_{\rm K}$ and $g_{\rm M}$, respectively.
Both functions are expected to be of the order of unity, but they may
not be exactly equal; see Kleeorin et al.\ (2002).
In the following we combine the information from the kinetically and
magnetically driven runs to compute separately $\tau g_{\rm K}$ and
$\tau g_{\rm M}$, and assume
$\alpha_{\rm K}=\tau g_{\rm K}\tilde\alpha_{\rm K}$
and $\alpha_{\rm M}=\tau g_{\rm M}\tilde\alpha_{\rm M}$.
We thus have an equation with two unknowns,
\EQ
\alpha=\tau g_{\rm K}\tilde\alpha_{\rm K}+\tau g_{\rm M}\tilde\alpha_{\rm M},
\label{OneUnknown}
\EN
where the two unknowns are $\tau g_{\rm K}$ and $\tau g_{\rm M}$.

\begin{table*}[htb]\caption{
Summary of the measured rms velocity and normalized $\alpha$s
for kinetically and magnetically forced runs, together with
the resulting Strouhal numbers and their corresponding errors
obtained by taking the extrema from 5 separate averages over
1/5 of the full time series.
Values in parenthesis show departures from the trend
and should be regarded as uncertain.
For $k_{\rm f}=1.5$ the resolution varies between $64^3$ and $512^3$
meshpoints for $\eta=2\times10^{-3}$ and $2\times10^{-4}$, respectively,
while for $k_{\rm f}=5$ the resolution varies between $32^3$ and $256^3$
meshpoints for $\eta=5\times10^{-3}$ and $5\times10^{-4}$, respectively.
}\vspace{12pt}\centerline{\begin{tabular}{lcc|cccc|cccc|cccc}
$B_0$ & $\eta$ & $k_{\rm f}$ &
$u_{\rm rms}^{\rm(k)}$ & $a^{\rm(k)}$ &
$\tilde{a}_{\rm K}^{\rm(k)}$ & $\tilde{a}_{\rm M}^{\rm(k)}$ &
$u_{\rm rms}^{\rm(m)}$ & $a^{\rm(m)}$ &
$\tilde{a}_{\rm K}^{\rm(m)}$ & $\tilde{a}_{\rm M}^{\rm(m)}$ &
$\mbox{St}\,g_{\rm K}$ & $\mbox{St}\,g_{\rm M}$ &
$\delta\mbox{St}\,g_{\rm K}$ & $\delta\mbox{St}\,g_{\rm M}$ \\
\hline
0.01&$2\times10^{-3}$&1.5&$ 0.10$&$-0.261$&$-0.46$&$ 0.04$&$ 0.05$& (4.79)&$-0.11$&$ 1.44$&$ 0.89$& (3.40)&$ 0.07$& (1.29)\\
0.03&$2\times10^{-4}$&1.5&$ 0.09$&($-0.042$)&$-0.37$&$ 0.44$&$ 0.06$&$ 0.36$&$-0.12$&$ 2.04$&(0.35)&(0.20)& (0.09)& (0.04)\\
0.03&$5\times10^{-4}$&1.5&$ 0.09$&$-0.062$&$-0.37$&$ 0.42$&$ 0.06$&$ 0.88$&$-0.13$&$ 1.86$&$ 0.76$&$ 0.53$&$ 0.16$&$ 0.08$\\
0.03&$1\times10^{-3}$&1.5&$ 0.09$&$-0.099$&$-0.39$&$ 0.32$&$ 0.05$&$ 0.88$&$-0.13$&$ 1.31$&$ 0.86$&$ 0.76$&$ 0.11$&$ 0.23$\\
0.03&$2\times10^{-3}$&1.5&$ 0.09$&$-0.143$&$-0.42$&$ 0.24$&$ 0.05$&$ 0.74$&$-0.14$&$ 1.12$&$ 0.78$&$ 0.76$&$ 0.01$&$ 0.21$\\
0.06&$1\times10^{-3}$&1.5&$ 0.09$&$-0.030$&$-0.40$&$ 0.36$&$ 0.06$&$ 0.23$&$-0.24$&$ 0.61$&$ 0.65$&$ 0.63$&$ 0.06$&$ 0.09$\\
0.06&$2\times10^{-3}$&1.5&$ 0.08$&$-0.054$&$-0.40$&$ 0.35$&$ 0.05$&$ 0.22$&$-0.24$&$ 0.58$&$ 0.71$&$ 0.67$&$ 0.03$&$ 0.10$\\
0.10&$2\times10^{-4}$&1.5&$ 0.12$&$-0.004$&$-0.42$&$ 0.25$&$ 0.09$&$ 0.07$&$-0.24$&$ 0.43$&$ 0.15$&$ 0.24$&$ 0.07$&$ 0.13$\\
0.10&$5\times10^{-4}$&1.5&$ 0.10$&$-0.004$&$-0.40$&$ 0.32$&$ 0.07$&$ 0.08$&$-0.30$&$ 0.48$&$ 0.28$&$ 0.34$&$ 0.03$&$ 0.04$\\
0.10&$1\times10^{-3}$&1.5&$ 0.10$&$-0.010$&$-0.43$&$ 0.32$&$ 0.07$&$ 0.08$&$-0.29$&$ 0.46$&$ 0.27$&$ 0.34$&$ 0.01$&$ 0.04$\\
0.10&$2\times10^{-3}$&1.5&$ 0.09$&$-0.019$&$-0.43$&$ 0.30$&$ 0.06$&$ 0.07$&$-0.28$&$ 0.45$&$ 0.28$&$ 0.34$&$ 0.01$&$ 0.05$\\
0.14&$2\times10^{-3}$&1.5&$ 0.10$&$-0.009$&$-0.43$&$ 0.26$&$ 0.06$&$ 0.04$&$-0.28$&$ 0.45$&$ 0.11$&$ 0.15$&$ 0.00$&$ 0.02$\\
0.20&$2\times10^{-3}$&1.5&$ 0.11$&$-0.004$&$-0.43$&$ 0.21$&$ 0.06$&$ 0.02$&$-0.27$&$ 0.43$&$ 0.04$&$ 0.06$&$ 0.00$&$ 0.01$\\
0.30&$2\times10^{-3}$&1.5&$ 0.12$&$-0.002$&$-0.42$&$ 0.18$&$ 0.06$&$ 0.01$&$-0.24$&$ 0.41$&$ 0.01$&$ 0.02$&$ 0.00$&$ 0.00$\\
0.06&$5\times10^{-4}$&5  &$ 0.16$&$-0.080$&$-0.31$&$ 0.25$&$ 0.15$&$ 0.10$&$-0.27$&$ 0.89$&$ 0.46$&$ 0.25$&$ 0.00$&$ 0.01$\\
0.06&$1\times10^{-3}$&5  &$ 0.16$&$-0.121$&$-0.32$&$ 0.20$&$ 0.14$& (0.01)&$-0.12$&$ 2.03$&$ 0.39$& (0.03)&$ 0.02$& (0.00)\\
0.06&$2\times10^{-3}$&5  &$ 0.15$&$-0.172$&$-0.49$&$ 0.22$&$ 0.06$&$ 0.34$&$-0.16$&$ 0.52$&$ 0.75$&$ 0.89$&$ 0.09$&$ 0.25$\\
0.06&$5\times10^{-3}$&5  &$ 0.13$&$-0.215$&$-0.41$&$ 0.10$&$ 0.08$&$ 0.54$&$-0.18$&$ 0.81$&$ 0.74$&$ 0.83$&$ 0.12$&$ 0.02$\\
0.10&$5\times10^{-4}$&5  &$ 0.16$&$-0.035$&$-0.32$&$ 0.30$&$ 0.15$&$ 0.36$&$-0.20$&$ 0.72$&$ 0.79$&$ 0.72$&$ 0.03$&$ 0.03$\\
0.10&$1\times10^{-3}$&5  &$ 0.15$&$-0.058$&$-0.34$&$ 0.27$&$ 0.13$&$ 0.35$&$-0.21$&$ 0.70$&$ 0.76$&$ 0.72$&$ 0.04$&$ 0.06$\\
0.10&$2\times10^{-3}$&5  &$ 0.14$&$-0.091$&$-0.36$&$ 0.25$&$ 0.11$&$ 0.34$&$-0.22$&$ 0.72$&$ 0.73$&$ 0.70$&$ 0.01$&$ 0.06$\\
0.10&$5\times10^{-3}$&5  &$ 0.12$&$-0.131$&$-0.41$&$ 0.18$&$ 0.08$&$ 0.31$&$-0.24$&$ 0.75$&$ 0.59$&$ 0.60$&$ 0.03$&$ 0.08$\\
\label{Ttau}\end{tabular}}\end{table*}

In order to obtain a second equation, we can consider the simulations for
kinetically and magnetically driven turbulence as two independent measurement.
This leads to two separate measurements of
$\alpha$, $\tilde\alpha_{\rm K}$, and $\tilde\alpha_{\rm M}$,
distinguished by superscripts (k) and (m), respectively.
In this way, \Eq{OneUnknown} becomes a matrix equation
\EQ
\pmatrix{\alpha^{\rm(k)}\cr\alpha^{(m)}}=
\pmatrix{
\tilde\alpha_{\rm K}^{\rm(k)}&\tilde\alpha_{\rm M}^{\rm(k)}\cr
\tilde\alpha_{\rm K}^{(m)}&\tilde\alpha_{\rm M}^{(m)}}
\pmatrix{\tau g_{\rm K}\cr\tau g_{\rm M}},
\label{TwoUnknowns}
\EN
which can be solved for $\tau g_{\rm K}$ and $\tau g_{\rm M}$.
However, since the root mean square velocities,
$u_{\rm rms}^{\rm(k)}$ and $u_{\rm rms}^{\rm(m)}$,
are different in the kinetically and magnetically driven cases,
we non-dimensionalize each measurement independently, so we define
\EQ
a^{\rm(k,m)}=\alpha^{\rm(k,m)}/u_{\rm rms}^{\rm(k,m)},
\EN
\EQ
\tilde{a}^{\rm(k,m)}_{\rm K,M}=
\tilde\alpha^{\rm(k,m)}_{\rm K,M}/
\left[k_{\rm f}u_{\rm rms}u_{\rm rms}^{\rm(k,m)}\right],
\EN
\EQ
\mbox{St}\,g_{\rm K,M}=u_{\rm rms}k_{\rm f}\tau g_{\rm K,M},
\EN
where $u_{\rm rms}=[u_{\rm rms}^{\rm(k)}u_{\rm rms}^{\rm(m)}]^{1/2}$
is the geometrical mean of the rms velocities for kinetically and
magnetically driven runs.
Thus, the matrix equation that we actually solve is
\EQ
\pmatrix{a^{\rm(k)}\cr a^{(m)}}=
\pmatrix{
\tilde{a}_{\rm K}^{\rm(k)}&\tilde{a}_{\rm M}^{\rm(k)}\cr
\tilde{a}_{\rm K}^{(m)}&\tilde{a}_{\rm M}^{(m)}}
\pmatrix{\mbox{St}\,g_{\rm K}\cr\mbox{St}\,g_{\rm M}}.
\label{TwoUnknowns}
\EN
The result is shown in \Fig{ptau}.
Note that, unlike the previous case, $\tau g_{\rm K}$ and $\tau g_{\rm M}$
are now always positive.
In some cases the results are quite similar to the previously determined
values of $\tau$, but in the case with $B_0=0.03$ and $k_{\rm f}=1.5$,
where $\tau$ became negative (not seen in the logarithmic representation
in \Fig{ptau1}), both $\tau g_{\rm K}$ and $\tau g_{\rm M}$
are now positive, and approximately constant for $R_{\rm m}\leq100$.
However, for larger values of $R_{\rm m}$, both values
may decrease, although it should be noted that the error
bars are also larger and the accuracy of the error itself may not be
reliable either.
Thus, it is not yet clear that the decline seen for the largest value of
$R_{\rm m}$ is indeed real.

\begin{figure}[t!]\begin{center}
\includegraphics[width=\columnwidth]{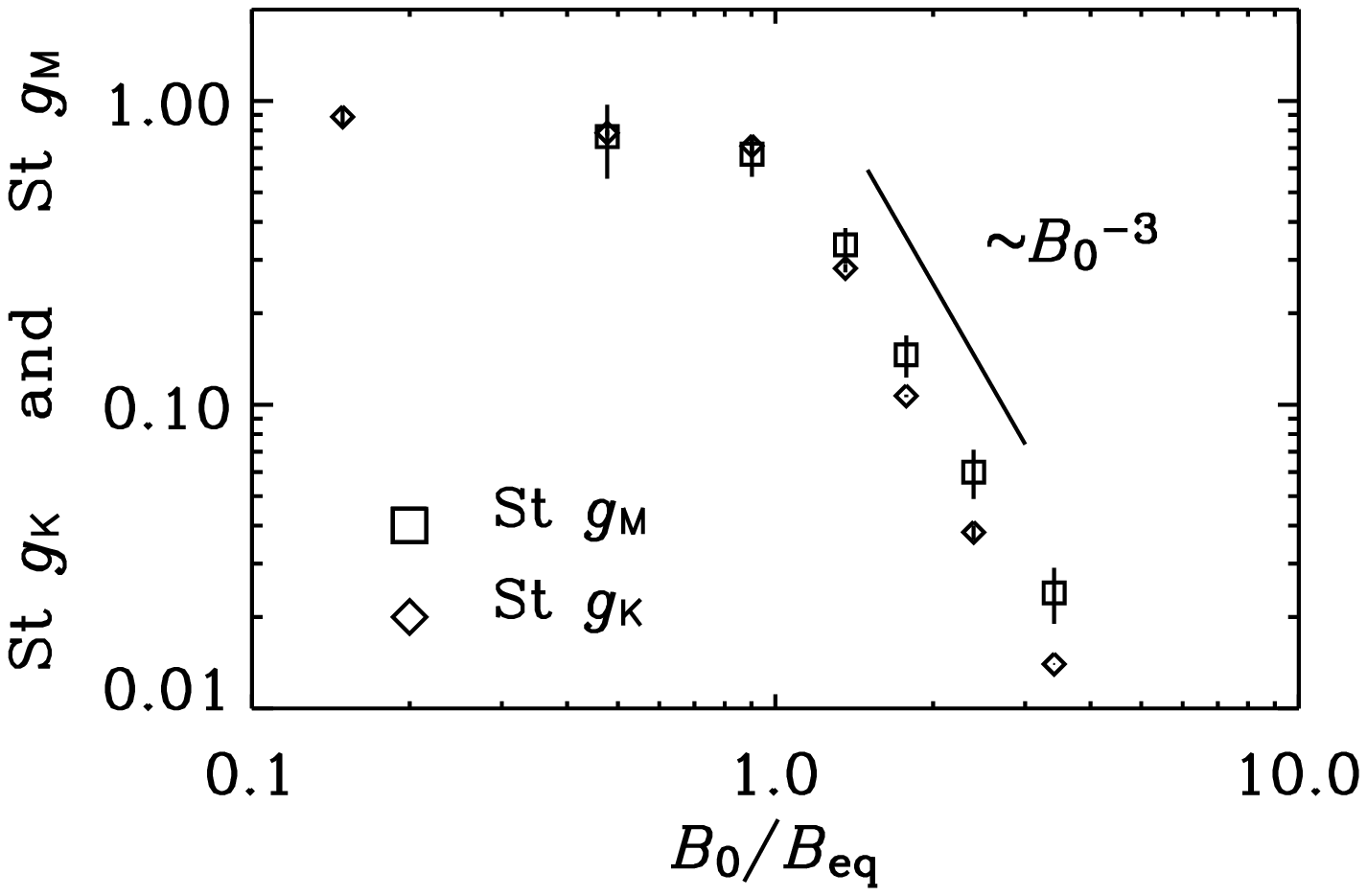}
\end{center}\caption[]{
Magnetic and kinetic Strouhal numbers as a function of $B_0/B_{\rm eq}$
for $\eta=2\times10^{-3}$ and  $k_{\rm f}=1.5$.
Kinetically and mag\-ne\-ti\-cally forced runs have been used to
calculate separately $g_{\rm K}\neq g_{\rm M}$.
}\label{ptau_vs_B0}\end{figure}

When the field strength is increased, both $\tau\tilde\alpha_{\rm K}$ and
$\tau\tilde\alpha_{\rm M}$ are quenched uniformly, i.e.\ approximately
independently of $R_{\rm m}$.
This becomes clear when comparing the first two panels of \Fig{ptau},
or by inspecting the data of \Tab{Ttau}.
For a fixed value of $R_{\rm m}$ (around 50) the quenching with
increasing field strength is shown in \Fig{ptau_vs_B0}.
This quenching, which sets in once $B_0/B_{\rm eq}$ is of the order of unity,
is rather strong (proportional to $B_0^{-3}$).
Similarly strong quenching has previously been found for
$\alpha$ if the contribution from $\overline{\jj\cdot\bb}$
is ignored (Moffatt 1972, R\"udiger 1974, R\"udiger \& Kitchatinov 1993).

\section{Energy and helicity spectra}

\begin{figure}[t!]\begin{center}
\includegraphics[width=\columnwidth]{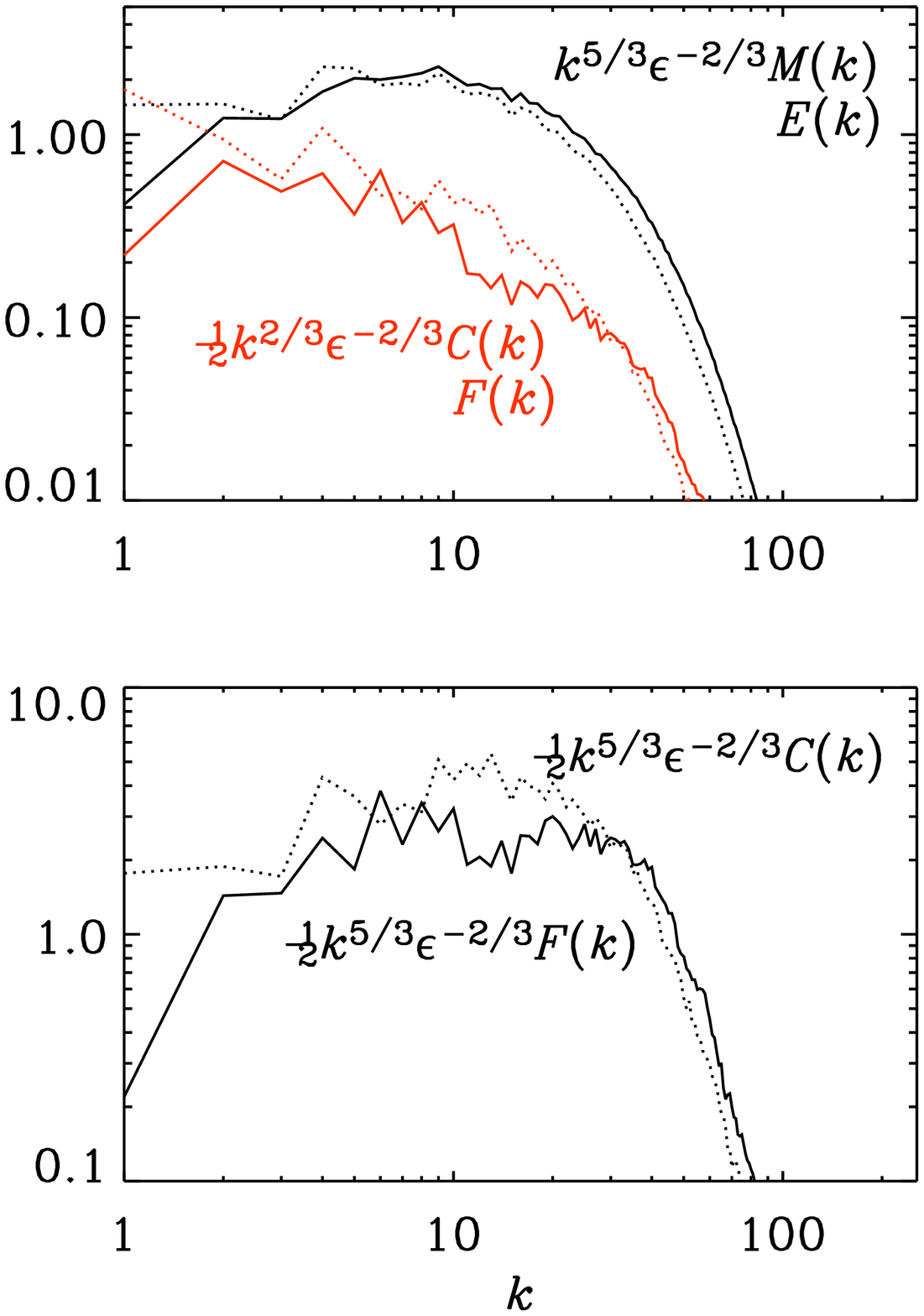}
\end{center}\caption[]{
Compensated shell-integrated power spectra of magnetic energy $M(k)$
and kinetic energy $E(k)$ (the upper solid and dotted lines in the
upper panel), and current helicity $C(k)$ and kinetic helicity $F(k)$
(the lower solid and dotted red or gray lines in the upper panel).
The energy spectra spectra a made dimensionless by scaling with
$\epsilon^{-2/3}$, where $\epsilon$ is the total energy dissipation rate.
The two helicity spectra are scaled such that,
if the fields were nearly fully helical at all scales, they would
be close to the corresponding energy spectra, which is not the case.
Instead, both energy and helicity spectra show a short subrange where
they are best fitted with a $k^{-5/3}$ power law.
The helicity spectra, compensated by $k^{5/3}$, are shown in the lower panel.
The turbulence is kinetically forced, with imposed field $B_0=0.1$, using
$512^3$ meshpoints.
}\label{ppowerhelH512b3}\end{figure}

In \Figs{ppowerhelH512b3}{ppowerhelM512b3} we present,
both for the kinetically and magnetically driven cases, shell-integrated
spectra of magnetic and kinetic energies, $M(k)$ and $E(k)$, respectively,
as well as current and kinetic helicities, $C(k)$ and $F(k)$, respectively.
They are normalized such that $\int M(k)\dd k=\half\bra{\BB^2}$,
$\int E(k)\dd k=\half\bra{\uu^2}$, $\int C(k)\dd k=\bra{\JJ\cdot\BB}$,
and $\int F(k)\dd k=\bra{\oo\cdot\uu}$.
The two helicity spectra are subject to a realizability condition,
$|C(k)|\leq2kM(k)$ and $|F(k)|\leq2kE(k)$ (Moffatt 1978).
If the bound is nearly saturated, we say that the field is fully helical.
It turns out that for $k>k_{\rm f}$ we have $|C(k)|\ll2kM(k)$
and $|F(k)|\ll2kE(k)$, so magnetic and velocity fields are
in fact not fully helical at small scales.

\begin{figure}[t!]\begin{center}
\includegraphics[width=\columnwidth]{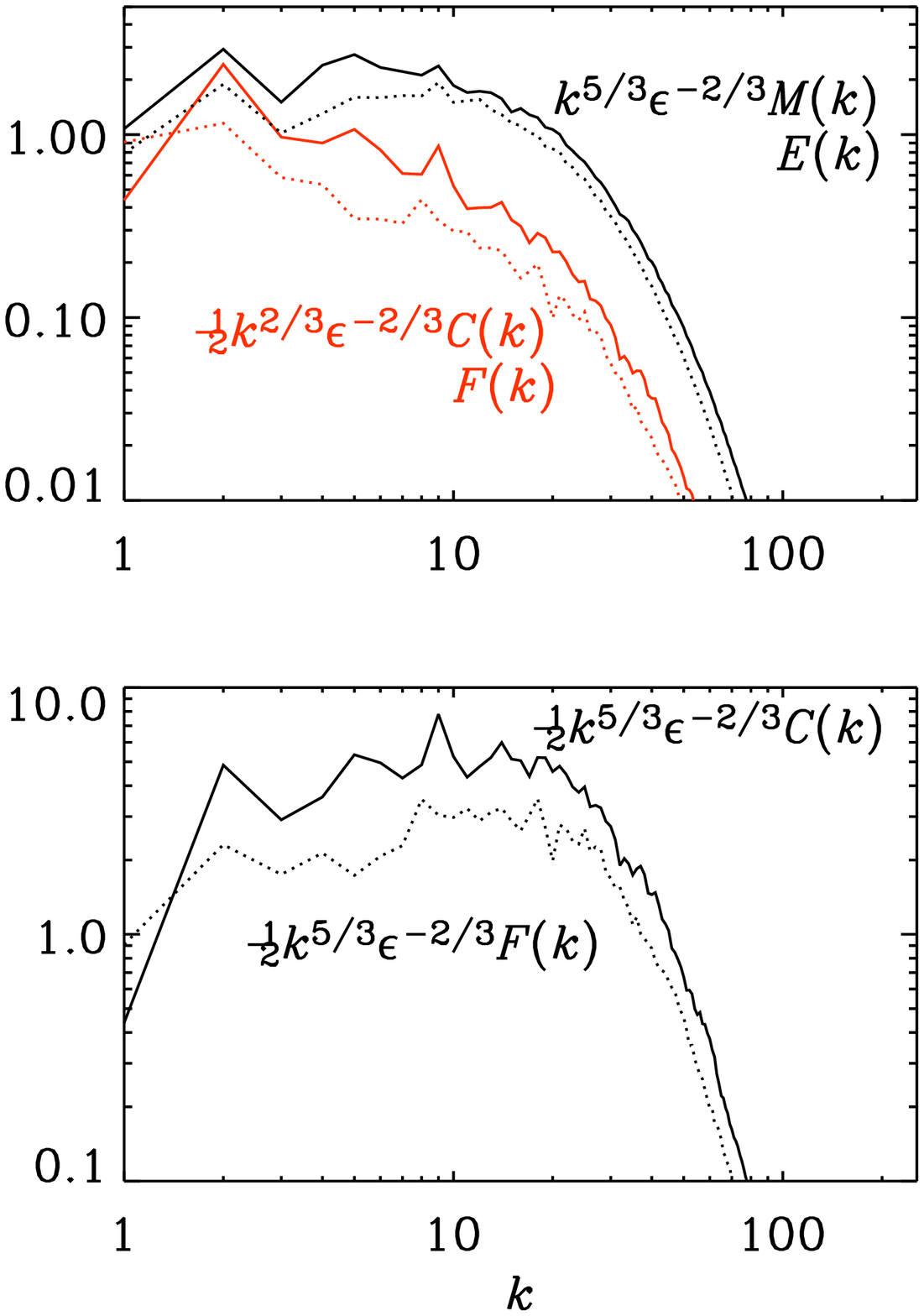}
\end{center}\caption[]{
Same as \Fig{ppowerhelH512b3}, but for the
magnetically forced case with $B_0=0.1$, $512^3$ meshpoints.
}\label{ppowerhelM512b3}\end{figure}

Since both $C(k)$ and $F(k)$ seem to scale like $k^{-5/3}$ in the inertial
range, both helicities are governed by contributions from large scales.
This implies in particular that the small scale current helicity,
$\bra{\jj\cdot\bb}$, and hence also $\tilde{\alpha}_{\rm M}$, should not
depend on the length of the inertial range and hence should be
independent of $R_{\rm m}$.
If this were not the case, i.e.\ if the magnetic field were fully helical
with a $k^{-2/3}$ current helicity spectrum, $\tilde{\alpha}_{\rm K}$ would be
asymptotically proportional to $R_{\rm m}^{1/4}$ (Blackman \& Brandenburg 2002).
Our results suggest that this is not the case here.
This seems to be in conflict with earlier results for helical dynamos
(Brandenburg 2001), but those results were at 4 times smaller resolution
($120^3$ instead of $512^3$ meshpoints)
and the forcing was at 3 times smaller scale, so
the effective length of the inertial range was 12 times smaller and
hence completely absent.
What remains true, however, is that the field is nearly fully helical
at the energy carrying scale, i.e.\ at wavenumber $k_{\rm f}$.

\section{Kinetically vs magnetically forced turbulence}

\begin{figure*}[t!]\begin{center}
\includegraphics[width=.85\columnwidth]{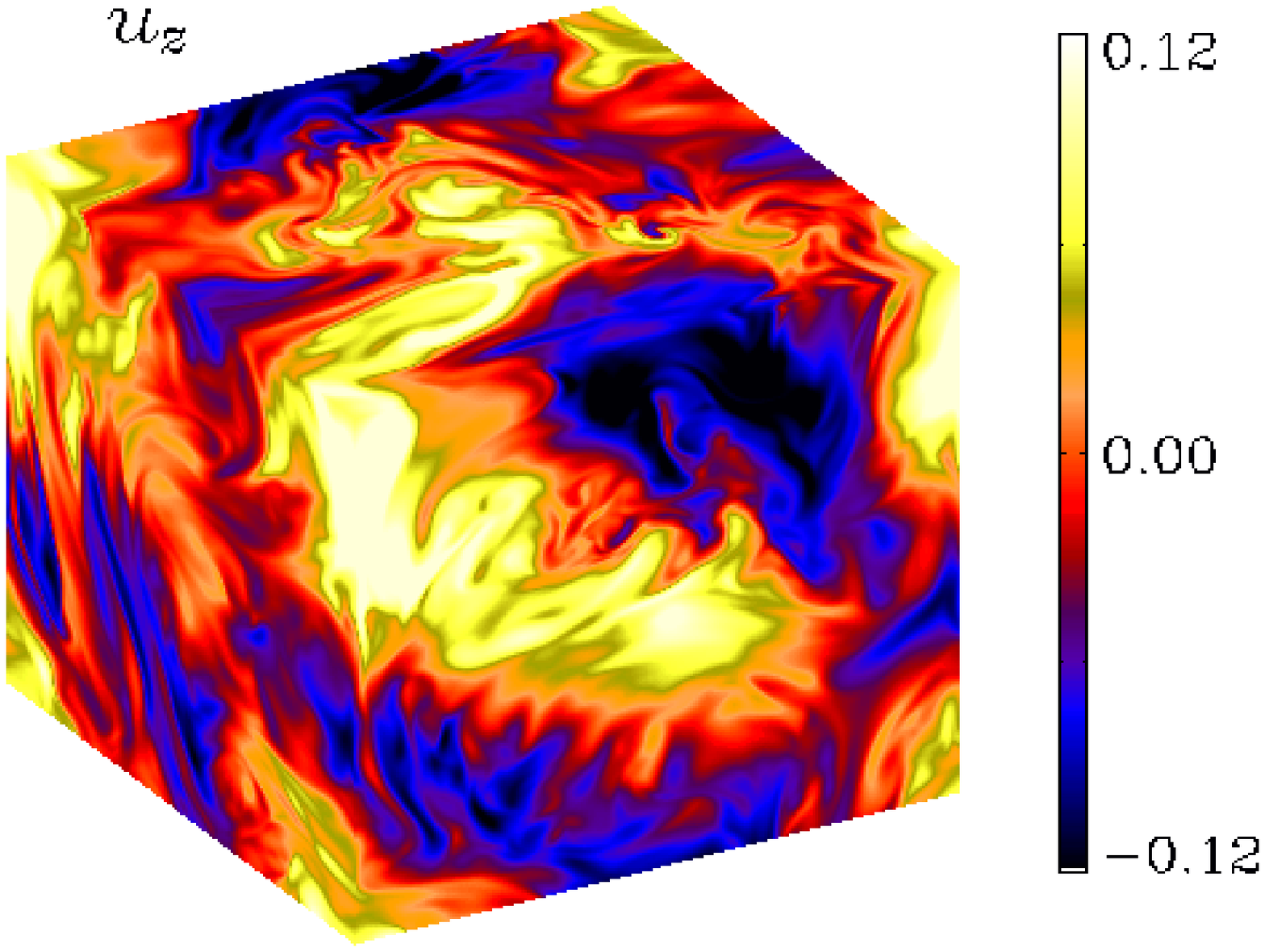}
\includegraphics[width=.85\columnwidth]{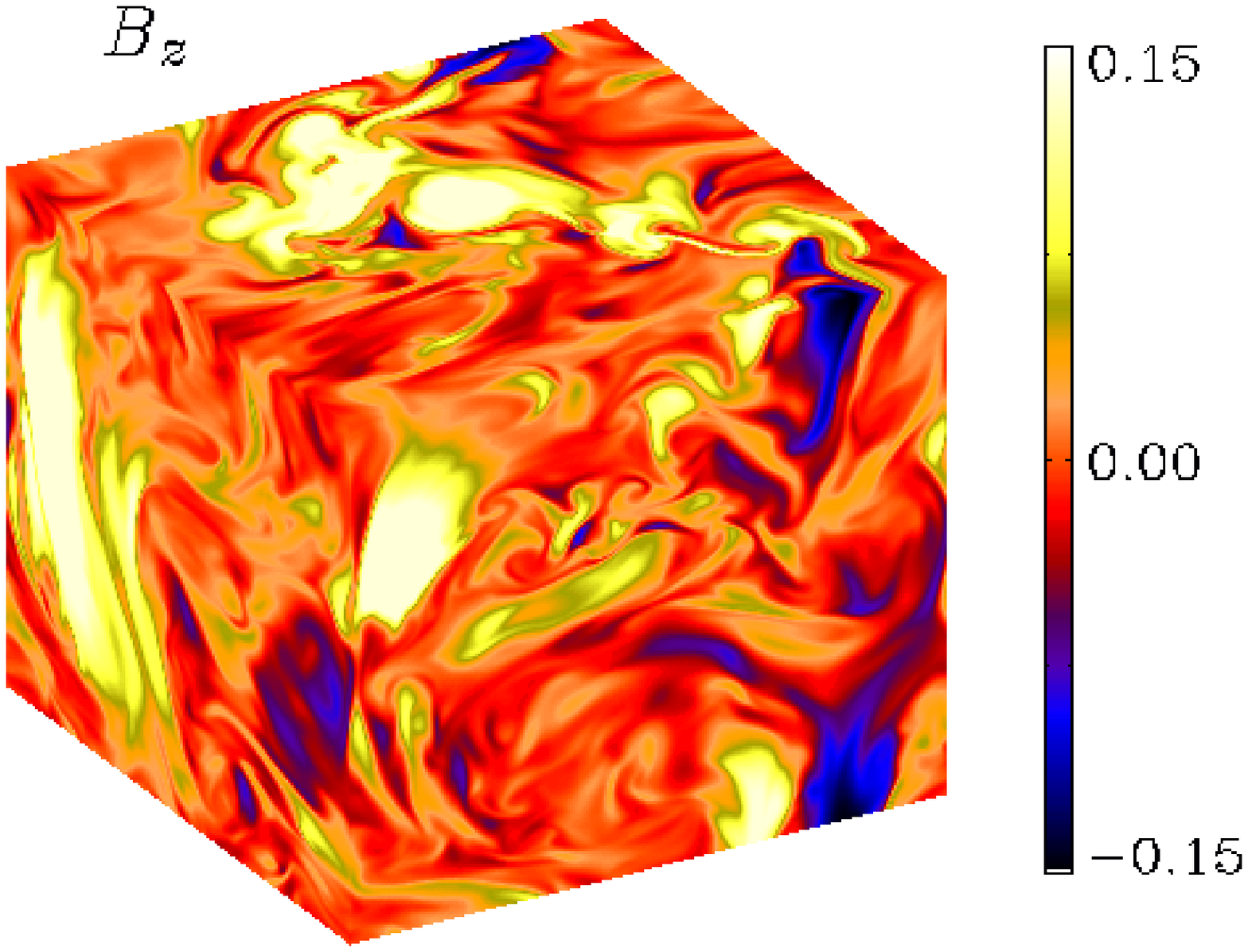}
\end{center}\caption[]{
Visualizations of $u_z$ (left) and $b_z$ (right) in the
kinetically forced case.
$B_0=0.03$, $\nu=\eta=2\times10^{-4}$, $k_{\rm f}=1.5$,
$512^3$ meshpoints.
}\label{H512e}\end{figure*}

In order to gain more insight into the nature of the turbulence
in kinetically and magnetically forced cases we show visualizations
of the field-aligned components of velocity and magnetic field,
$u_z$ and $B_z$, respectively.
In the kinetically driven case we see a large scale pattern in velocity,
consistent with power at the expected energy-carrying scale corresponding
to the forcing wavenumber $k_{\rm f}=1.5$; see \Fig{H512e}.
The magnetic field, on the other hand, is dominated by smaller scale
structures.
The strength of the imposed field is still less than the equipartition value
($B_0/B_{\rm eq}\approx0.4$) and there is no obvious anisotropy.
Although the velocity field is helical, there is no obvious large-scale
pattern in the magnetic field.
There are two possible reasons for this: lack of scale separation
(see Sect.~V of Haugen et al.\ 2004) and suppression of large scale
dynamo action by the large scale field (Montgomery et al.\ 2002,
Brandenburg \& Matthaeus 2004).

\begin{figure*}[t!]\begin{center}
\includegraphics[width=.85\columnwidth]{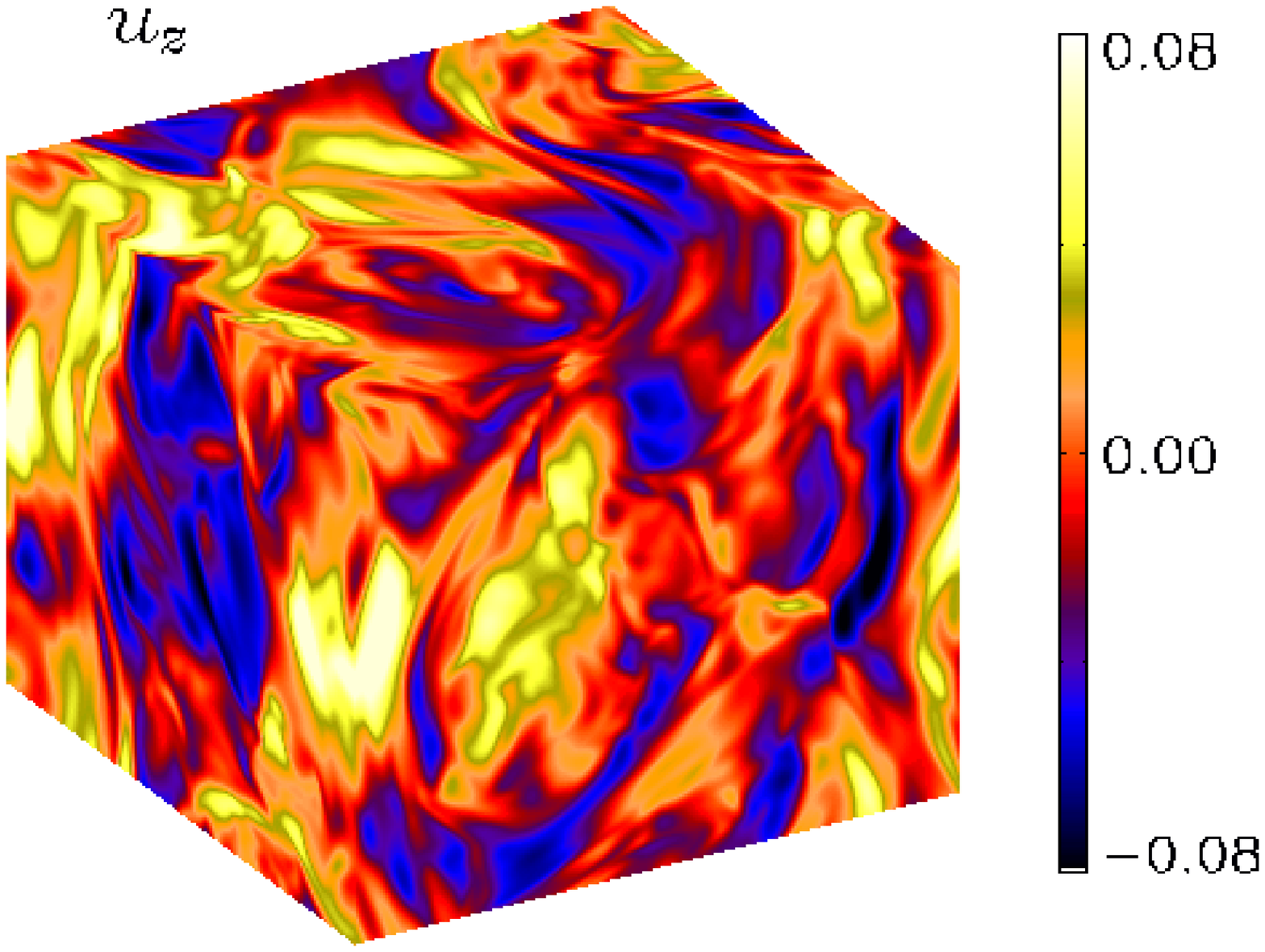}
\includegraphics[width=.85\columnwidth]{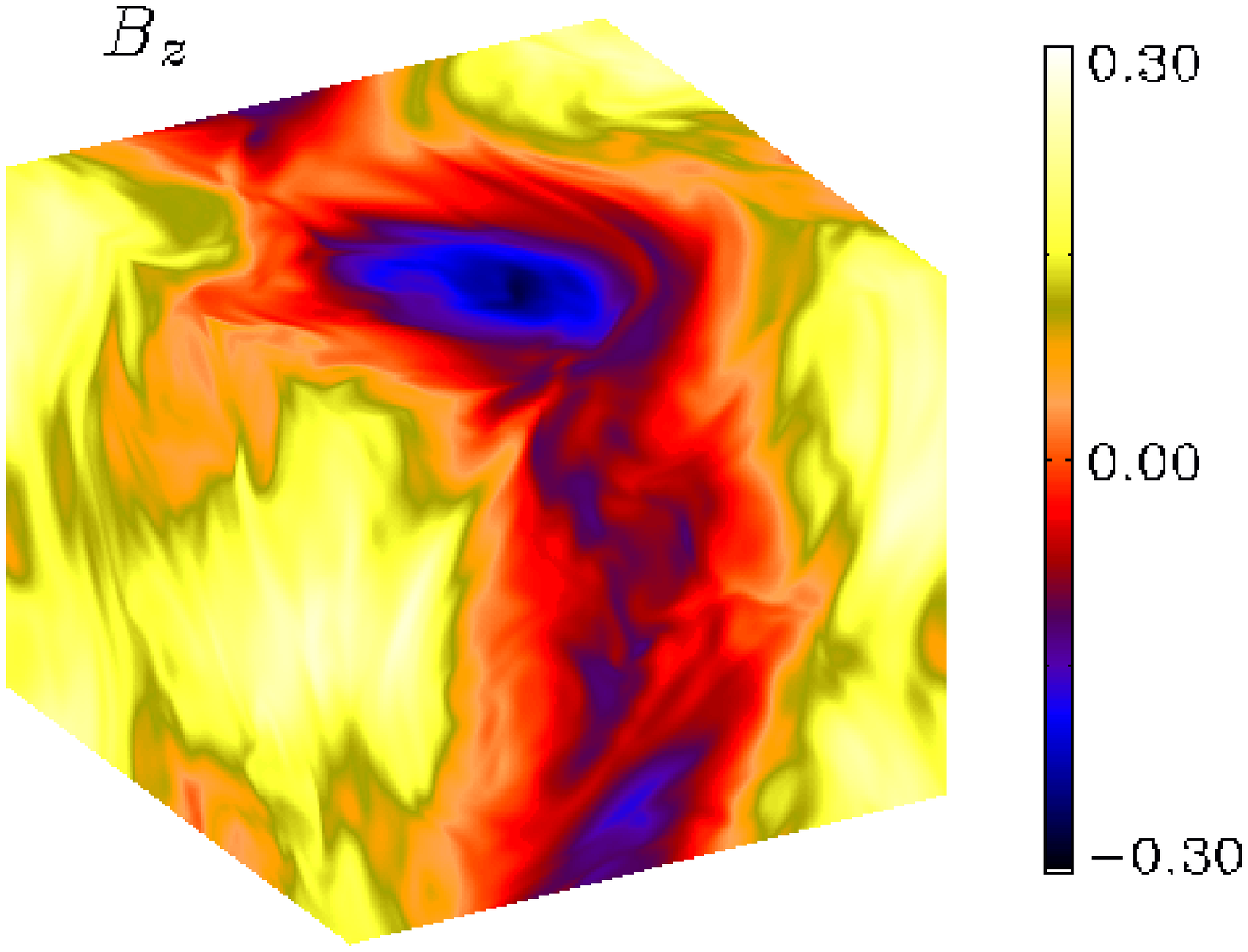}
\end{center}\caption[]{
Visualizations of $u_z$ (left) and $b_z$ (right) in the
magnetically forced case.
$B_0=0.03$, $\nu=\eta=2\times10^{-4}$, $k_{\rm f}=1.5$,
$512^3$ meshpoints.
}\label{M512e}\end{figure*}

\begin{figure*}[t!]\begin{center}
\includegraphics[width=.85\columnwidth]{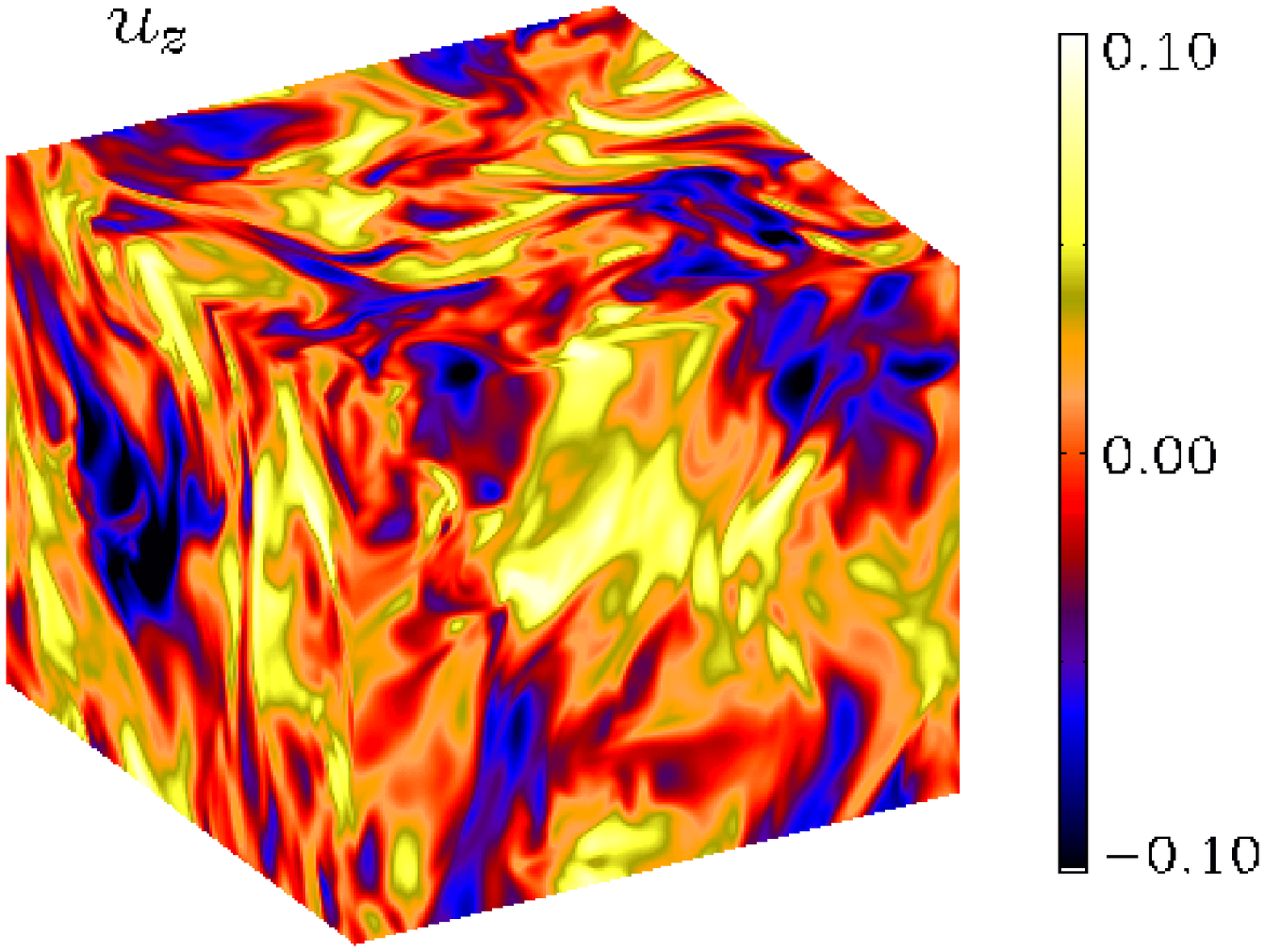}
\includegraphics[width=.85\columnwidth]{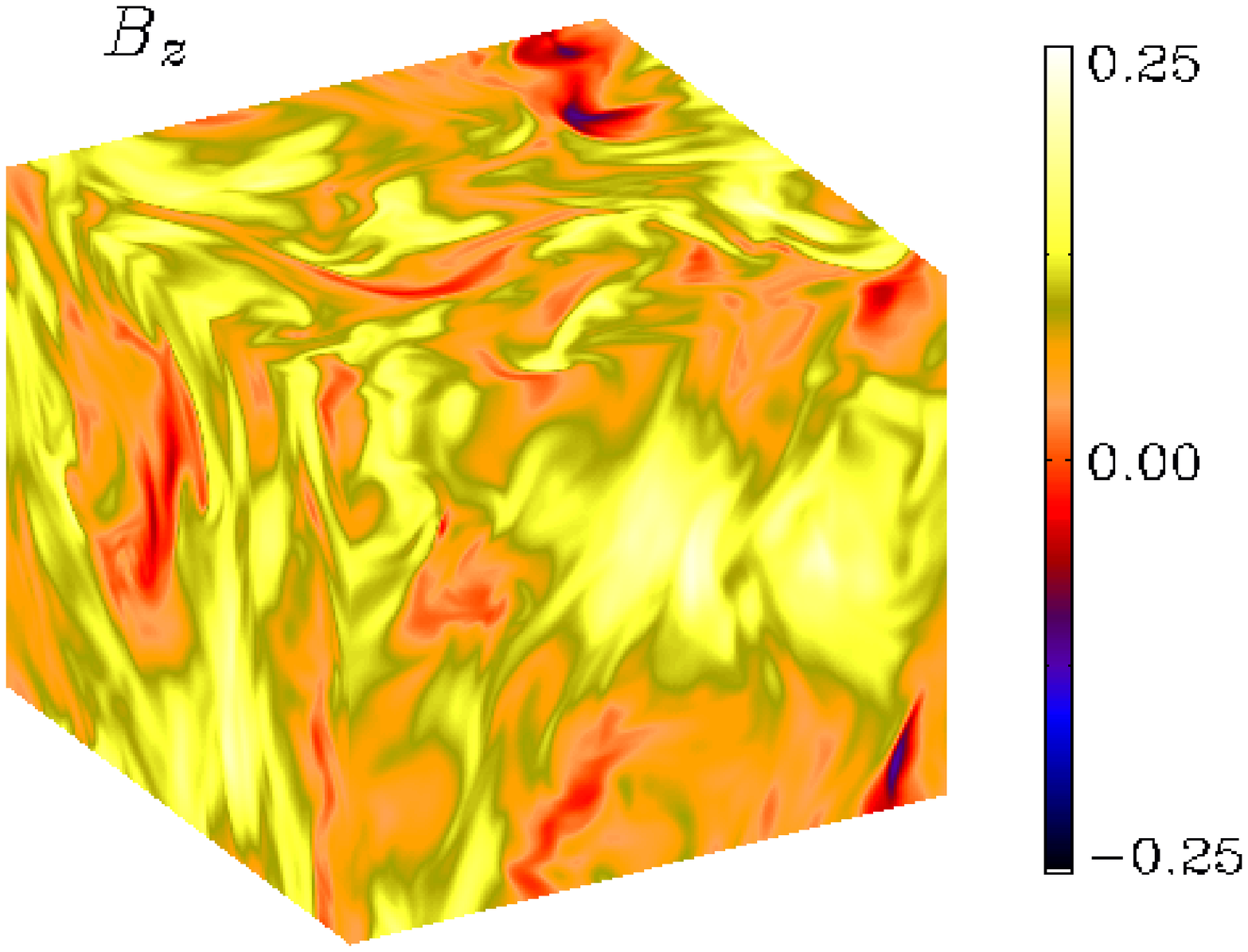}
\end{center}\caption[]{
Same as \Fig{M512e}, but for $B_0=0.1$.
There is no clear evidence for a large scale field.
}\label{M512b3}\end{figure*}

The magnetically driven case is quite different.
There seems to be a large scale pattern in the magnetic field (\Fig{M512e}).
Although there is no scale separation, it is possible to drive a larger
scale field in this case.
The main difference is that in the kinetically driven case,
dynamo-generated large scale and small scale fields have opposite sign
of current helicity (Brandenburg 2001).
In the absence of sufficient scale separation this leads to cancelation.
In the magnetically driven case the current helicity has the same sign
at all scales, so there is no cancellation and therefore the build-up
of a large scale field is possible.
However, when the field is too strong ($B_0=0.1$, corresponding to
$B_0/B_{\rm eq}\approx1$) no large scale pattern in the magnetic field
develops; see \Fig{M512b3}.

\section{Conclusions}

Although a closure hypothesis of the form $\meanTT=-\meanEMF/\tau$
has a plausible physical interpretation as a relaxation term,
(Blackman \& Field 2002), it is nevertheless
quite crude and lacks a rigorous physical basis.
Nevertheless, on empirical grounds this closure assumption is quite
appealing.
First of all, unlike the first order smoothing approximation, MTA provides
a natural and convincing explanation for the $\overline{\jj\cdot\bb}$
correction term to the $\alpha$ effect in \Eq{alpiso}.
This is indeed a key ingredient to the dynamical quenching model
(Kleeorin \& Ruzmaikin 1982), where the evolution of
$\overline{\jj\cdot\bb}$ is obtained by solving the magnetic helicity
equation for the small scale field.
The dynamical evolution of the $\overline{\jj\cdot\bb}$ term
provides a good model for the resistively limited saturation
(Field \& Blackman 2002, Blackman \& Brandenburg 2002, Subramanian 2002),
which is seen in helically forced simulations in periodic boxes
(Brandenburg 2001, Mininni et al.\ 2005).
Another appealing property of MTA is that the additional time derivative,
$\partial\meanEMF/\partial t$ in \Eq{dEMFdt}, restores causality
and prevents infinitely fast signal propagation.
In the case of turbulent passive scalar diffusion, one can easily see
that this term turns the heat equation into a damped wave equation,
where the maximum signal speed is the rms velocity of the turbulence
(Brandenburg et al.\ 2004).

One of the striking features of the dynamical quenching model is that
the resistively limited saturation and the resistive quenching of the
effective $\alpha$ can already be explained without invoking any explicit
$R_{\rm m}$ dependence of $\tau$.
Instead, the $R_{\rm m}$ dependence of the steady state values of
$\alpha$ emerges solely as a
result of magnetic helicity conservation and might hence be alleviated
if there is a helicity flux that offsets magnetic helicity conservation
(Blackman \& Field 2000, Kleeorin et al.\ 2000, 2002, Vishniac \& Cho 2001,
Subramanian \& Brandenburg 2004, Brandenburg \& Sandin 2004, Brandenburg 2005,
Brandenburg \& Subramanian 2005b).
However, there is no compulsory reason that there might not be
an explicit $R_{\rm m}$ dependence of $\tau$ after all.
Our present simulations begin to shed some light on this question.
However, the results are still not as clear cut as one would like
them to be.
Looking at \Fig{ptau}, it is evident that for $R_{\rm m}\leq100$, St
is independent of $R_{\rm m}$ and depends only on $\BB_0$; cf.\ the
uniform decrease of St between the first and second panels of \Fig{ptau}.
However, for $R_{\rm m}>100$, St may actually decrease with $R_{\rm m}$,
although the error bars also increase.
If this decrease is indeed real, it might indicate that the closure
hypothesis adopted here is too simplistic.

We should emphasize that, although we have in this work focused on
the $\alpha$ effect, this is not the only effect that produces large
scale dynamo action.
An example is the shear--current effect of
Rogachevskii \& Kleeorin (2003, 2004).
Here the electromotive force has a component in the direction
of $-\meanWW\times\meanJJ$, where $\meanWW=\nab\times\meanUU$ is the
vorticity of the mean flow.
(In the present work this effect is of course absent because there
is no mean flow and our mean fields are defined as volume averages,
so $\meanJJ=0$.)
Our work is relevant even in this nonhelical case because, like the
$\alpha$ effect, the $\meanWW\times\meanJJ$ effect produces mean fields
that are helical.
Depending on the amount of magnetic helicity losses from open
boundaries, there will be a tendency to limit the production of net
magnetic helicity within the domain, so small scale current helicity
of opposite sign must be produced at the same time as large scale
field is generated.
The presence of small scale current helicity implies that there must
be a finite $\alpha_{\rm M}$.
This is an example where there can be an $\alpha_{\rm M}$ term
even though $\alpha_{\rm K}=0$.
Related examples where $\alpha_{\rm K}=0$, but $\alpha_{\rm M}\neq0$,
are the reversed field pinch (e.g.\ Ho et al.\ 1989),
and the so-called selective decay (e.g.\ Montgomery et al.\ 1978),
as described in Blackman \& Brandenburg (2002) and Yousef et al.\ (2003).
Another example where $\alpha_{\rm K}=0$, but $\alpha_{\rm M}\neq0$,
is the Vishniac \& Cho (2001) mechanism whereby dynamo action is
accomplished through the current helicity flux that drives $\alpha_{\rm M}$;
see Brandenburg \& Subramanian (2005b) for model calculations showing that
this effect produces saturation field strengths independent of the
magnetic Reynolds number when the flux exceeds a certain threshold.

In the present paper we have only discussed the case where the
helicity of the turbulence is driven explicitly.
Although this is a common assumption that allows progress to be made in
that the flow can be taken to be statistically 
isotropic, it should be noted that in
astrophysical applications helicity is usually introduced by the
interaction between stratification and rotation (Krause \& R\"adler 1980).
To calculate this effect, MTA has to be used once more
(e.g.\ R\"adler et al.\ 2003), which then renders the $\alpha$ effect
proportional to $\tau^2$.
Nevertheless, as is emphasized in Brandenburg \& Subramanian (2005a),
also in this case the $\alpha$ effect is attenuated by the current
helicity contribution from the small scale field.

Earlier simulations of helical dynamos (no imposed field) suggested that
the field is bihelical, i.e.\ with one sign of helicity at large
scales, corresponding to wavenumber $k_1$, and the opposite sign at
small scales, corresponding to wavenumber $k_{\rm f}$ (Brandenburg 2001).
If the field were to continue to be nearly fully helical even at smaller
scales, this would have implied that $\tilde\alpha_{\rm M}$ would be
attenuated by a $R_{\rm m}^{1/4}$ factor (Blackman \& Brandenburg 2002).
Our present results suggest that this is not the case.
In agreement with both closure calculations (Andr\'e \& Lesieur 1977)
and direct numerical simulations (Borue \& Orszag 1997) we find
approximate $k^{-5/3}$ scaling for the kinetic helicity
(see also Ditlevsen \& Giuliani 2001).
Again, this implies that the velocity field is not fully helical beyond
wavenumber $k_{\rm f}$, so $\tilde\alpha_{\rm K}$ should also be
independent of $R_{\rm m}$.
This is indeed the case, except for very small values of $R_{\rm m}$.

There are various ways of extending the present studies of turbulent
transport to other fields.
An obvious possibility is turbulent viscosity and the $\Lambda$ effect
for modeling the Reynolds stress tensor (K\"apyl\"a et al.\ 2005).
For example, one would expect the relaxation time to depend on the
angular velocity (Kitchatinov \& R\"udiger 1993).
This dependence could be measured in a similar way as the dependence
on the magnetic field studied in the present paper.
Another related application is the study of passive scalar transport
in the present of rotation.
Both aspects are quite essential for many astrophysical applications.

\acknowledgements
We thank Eric G.\ Blackman for the many discussions we had on the subject,
and G\"unther R\"udiger for comments on the manuscript.
We also thank the organizers of the programs on ``Astrophysical Discs''
in Aspen (USA), ``Magnetohydrodynamics of Stellar Interiors'' at the
Isaac Newton Institute in Cambridge (UK), and ``Grand Challenge Problems
in Computational Astrophysics'' at the Institute for Pure and Applied
Mathematics at the University of California in Los Angeles, for providing
a stimulating environment.
The Danish Center for Scientific Computing is acknowledged for providing
time on the Linux cluster in Odense.

\appendix
\section{The forcing function}
\label{ForcingFunction}

For completeness we specify here the forcing function used in the
present paper\footnote{This forcing function was also used by
Brandenburg (2001), but in his Eq.~(5) the factor 2 in the denominator
should have been replaced by $\sqrt{2}$ for a proper normalization.}.
It is defined as
\EQ
\ff(\xx,t)={\rm Re}\{N\ff_{\kk(t)}\exp[\ii\kk(t)\cdot\xx+\ii\phi(t)]\},
\EN
where $\xx$ is the position vector.
The wavevector $\kk(t)$ and the random phase
$-\pi<\phi(t)\le\pi$ change at every time step, so $\ff(\xx,t)$ is
$\delta$-correlated in time.
For the time-integrated forcing function to be independent
of the length of the time step $\delta t$, the normalization factor $N$
has to be proportional to $\delta t^{-1/2}$.
On dimensional grounds it is chosen to be
$N=f_0 c_{\rm s}(|\kk|c_{\rm s}/\delta t)^{1/2}$, where $f_0$ is a
nondimensional forcing amplitude.
The value of the coefficient $f_0$ is chosen such that the maximum Mach
number stays below about 0.5; in practice this means $f_0=0.01\ldots0.05$,
depending on the average forcing wavenumber.
At each timestep we select randomly one of many possible wavevectors
in a certain range around a given forcing wavenumber.
The average wavenumber is referred to as $k_{\rm f}$.
Two different wavenumber intervals are considered: $1...2$ for
$k_{\rm f}=1.5$ and $4.5...5.5$ for $k_{\rm f}=5$.
We force the system with transverse helical waves,
\begin{equation}
\ff_{\kk}=\RRRR\cdot\ff_{\kk}^{\rm(nohel)}\quad\mbox{with}\quad
{\sf R}_{ij}={\delta_{ij}-\ii\sigma\epsilon_{ijk}\hat{k}_k
\over\sqrt{1+\sigma^2}},
\end{equation}
where $\sigma=1$ for positive helicity of the forcing function,
\EQ
\ff_{\kk}^{\rm(nohel)}=
\left(\kk\times\eee\right)/\sqrt{\kk^2-(\kk\cdot\eee)^2},
\label{nohel_forcing}
\EN
is a non-helical forcing function, and $\eee$ is an arbitrary unit vector
not aligned with $\kk$; note that $|\ff_{\kk}|^2=1$.

\section{Modified Keinigs relation}
\label{Keinigs}

The $R_{\rm m}$ dependence of $\alpha$ can generally be understood
as a direct consequence of magnetic helicity conservation
(Gruzinov \& Diamond 1994).
Dotting \Eq{bdot} with the vector potential $\aaaa$, where
$\bb=\nab\times\aaaa$, and adding the corresponding evolution of
$\dot{\aaaa}\cdot\bb$, averaging over a periodic volume, and using
$(\uu\times\BB_0)\times\bb=-(\uu\times\bb)\cdot\BB_0$ we obtain
\EQ
\half{\dd\over\dd t}\bra{\aaaa\cdot\bb}=-\bra{\uu\times\bb}\cdot\BB_0
-\eta\bra{\jj\cdot\bb}+\bra{\bb\cdot\ff_{\rm mag}}.
\EN
Using $\bra{\uu\times\bb}=\alpha\BB_0$, we find in the steady state
\EQ
\alpha=-\eta{\bra{\jj\cdot\bb}\over\BB_0^2}
+{\bra{\bb\cdot\ff_{\rm mag}}\over\BB_0^2}.
\EN
The first term, which was already obtained in an early paper by
Keinigs (1983), is manifestly $\eta$-dependent [see also
Brandenburg \& Matthaeus (2004) for a derivation of this term],
but the second term is not.
This explains the behavior seen in \Fig{palp_vs_Rm}.



\begin{thebibliography}{99}

\bibitem[]{}
Andr\'e, J.-C. \& Lesieur, M.\yjfm{1977}{81}{187}

\bibitem[]{}
Blackman, E. G., \& Field, G. F.\ymn{2000}{318}{724}

\bibitem[]{}
Blackman, E. G., \& Field, G. B.\yprl{2002}{89}{265007}

\bibitem[]{}
Blackman, E. G., \& Field, G. B.\ypf{2003}{15}{L73}

\bibitem[]{}
Blackman, E. G., \& Brandenburg, A.\yapj{2002}{579}{359}

\bibitem[]{}
Borue, V., \& Orszag, S. A.\ypre{1997}{55}{7005}

\bibitem[]{}
Brandenburg, A.\yapj{2001}{550}{824}

\bibitem[]{}
Brandenburg, A.\yapj{2005}{625}{539}

\bibitem[]{}
Brandenburg, A., \& Matthaeus, W. H.\ypre{2004}{69}{056407}

\bibitem[]{}
Brandenburg, A., \& Sandin, C.\yana{2004}{427}{13}

\bibitem[]{}
Brandenburg, A., \& Subramanian, K.\pprt{2005a}{{\sf astro-ph/0405052}}

\bibitem[]{}
Brandenburg, A., \& Subramanian, K.\yan{2005b}{326}{400}

\bibitem[]{}
Brandenburg, A., K\"apyl\"a, P., \& Mohammed, A.\ypf{2004}{16}{1020}

\bibitem[]{}
Cattaneo, F., \& Hughes, D. W.\ypre{1996}{54}{R4532}

\bibitem[]{}
Ditlevsen, P. D., \& Giuliani, P.\ypre{2001}{63}{036304}

\bibitem[]{}
Field, G. B., \& Blackman, E. G.\yapj{2002}{572}{685}

\bibitem[]{}
Gruzinov, A. V., \& Diamond, P. H.\yprl{1994}{72}{1651}

\bibitem[]{}
Haugen, N. E. L., Brandenburg, A., \& Dobler, W.\ypre{2004}{70}{016308}

\bibitem[]{}
Ho, Y. L., Prager, S. C., Schnack, D. D.\yprl{1989}{62}{1504}

\bibitem[]{}
K\"apyla, P. J., Korpi, M. J., Ossendrijver, M., \&
Tuominen, I.\yan{2005}{326}{186}

\bibitem[]{}
Keinigs, R. K.\ypf{1983}{26}{2558}

\bibitem[]{}
Kitchatinov, \& L. L., R\"udiger, G.\yana{1993}{276}{96}

\bibitem[]{}
Kitchatinov, L. L., R\"udiger, G., \& Pipin, V. V.\yan{1994}{315}{157}

\bibitem[]{}
Kleeorin, N. I., \& Ruzmaikin,
A. A.\yjour{1982}{Magne\-to\-hydro\-dynamics}{18}{116}

\bibitem[]{}
Kleeorin, N. I., Rogachevskii, I. V., Ruzmaikin, A. A.\yjetp{1990}{70}{878}

\bibitem[]{}
Kleeorin, N. I., Mond, M., \& Rogachevskii, I.\yana{1996}{307}{293}

\bibitem[]{}
Kleeorin, N. I, Moss, D., Rogachevskii, I., \& Sokoloff, D.\yana{2000}{361}{L5}

\bibitem[]{}
Kleeorin, N. I, Moss, D., Rogachevskii, I., \& Sokoloff, D.\yana{2002}{387}{453}

\bibitem[]{}
Krause, F., \& R\"adler, K.-H.\ybook{1980}
{Mean-Field Magneto\-hydro\-dy\-na\-mics and Dynamo Theory}
{Pergamon Press, Oxford}

\bibitem[]{}
Mininni, P. D., G\'omez, D. O., Mahajan, S. M.\yapj{2005}{619}{1019}

\bibitem[]{}
Moffatt, H. K.\yjfm{1972}{53}{385}

\bibitem[]{}
Moffatt, H. K.\ybook{1978}
{Magnetic Field Generation in Electrically Conducting Fluids}
{Cambridge University Press, Cambridge}

\bibitem[]{}
Montgomery, D., Turner, L., \& Vahala, G.\ypf{1978}{21}{757}

\bibitem[]{}
Montgomery, D., Matthaeus, W. H., Milano, L. J., \& Dmitruk,
P.\ypp{2002}{9}{1221}

\bibitem[]{}
Pouquet, A., Frisch, U., \& L\'eorat, J.\yjfm{1976}{77}{321}

\bibitem[]{}
R\"adler, K.-H., Kleeorin, N., \& Rogachevskii, I.\ygafd{2003}{97}{249}

\bibitem[]{}
Rogachevskii, I., \& Kleeorin, N.\ypre{2003}{68}{036301}

\bibitem[]{}
Rogachevskii, I., \& Kleeorin, N.\ypre{2004}{70}{046310}

\bibitem[]{}
R\"udiger, G.\yan{1974}{295}{275}

\bibitem[]{}
R\"udiger, G.\ybook{1989}{Differential rotation and stellar convection:
Sun and solar-type stars}{Gordon \& Breach, New York}

\bibitem[]{}
R\"udiger, G. \& Kitchatinov, L. L.\yana{1993}{269}{581}

\bibitem[]{}
Subramanian, K.\yjour{2002}{Bull.\ Astr.\ Soc.\ India}{30}{715}

\bibitem[]{}
Subramanian, K., \& Brandenburg, A.\yprl{2004}{93}{205001}

\bibitem[]{}
Vainshtein, S. I. \& Kitchatinov, L. L.\ygafd{1983}{24}{273}

\bibitem[]{}
Vishniac, E. T., \& Cho, J.\yapj{2001}{550}{752}

\bibitem[]{}
Yousef, T. A., Brandenburg, A., \& R\"udiger, G.\yana{2003}{411}{321}

\end{thebibliography}
\end{document}